%% file: main.tex
\pdfoutput=1
\documentclass[12pt,a4paper]{article}

\usepackage{ifthen} 
\usepackage{slashed}
\usepackage{comment}
\usepackage{multirow}
\newboolean{pdflatex}
\setboolean{pdflatex}{true} 
\usepackage{graphicx} 

\newboolean{articletitles}
\setboolean{articletitles}{true} 

\newboolean{uprightparticles}
\setboolean{uprightparticles}{false} 

\def\paperauthors{LHCb collaboration} 
\def\paperasciititle{Lambda to proton muon neutrino branching fraction measurement at LHCb with LHC Run-2 Data} 
\def\papertitle{Branching fraction measurement of the $\PLambda \rightarrow \proton \mun \neumb \ $ decay}
\def\paperkeywords{{High Energy Physics}, {LHCb}} 
\def\papercopyright{\the\year\ CERN for the benefit of the LHCb collaboration} 
\def\paperlicence{CC BY 4.0 licence}
\def\paperlicenceurl{https://creativecommons.org/licenses/by/4.0/}

\newif\ifEnableSectionTOCLinks
\EnableSectionTOCLinksfalse 

\input{preamble}
\usepackage{longtable} 

\begin{document}

\renewcommand{\thefootnote}{\fnsymbol{footnote}}
\setcounter{footnote}{1}

\input{title-LHCb-PAPER}


\renewcommand{\thefootnote}{\arabic{footnote}}
\setcounter{footnote}{0}


\cleardoublepage


\pagestyle{plain} 
\setcounter{page}{1}
\pagenumbering{arabic}


\input{body}

\input{acknowledgements}



\clearpage

\addcontentsline{toc}{section}{References}
\bibliographystyle{LHCb}
\bibliography{main,standard,LHCb-PAPER,LHCb-CONF,LHCb-DP,LHCb-TDR}

\newpage
\input{Authorship_LHCb-PAPER-2025-030}




\end{document}

%% file: preamble.tex
\usepackage[top=1in, bottom=1.25in, left=1in, right=1in]{geometry}

\usepackage{microtype}
\usepackage{lineno}  
\usepackage{xspace} 
\usepackage{caption} 

\usepackage{graphicx}  
\usepackage{color}
\usepackage{colortbl}
\graphicspath{{./figs/}} 

\usepackage{amsmath} 
\usepackage{amssymb}
\usepackage{amsfonts}
\usepackage{upgreek} 

\newcommand*\patchAmsMathEnvironmentForLineno[1]{%
\expandafter\let\csname old#1\expandafter\endcsname\csname #1\endcsname
\expandafter\let\csname oldend#1\expandafter\endcsname\csname
end#1\endcsname
 \renewenvironment{#1}%
   {\linenomath\csname old#1\endcsname}%
   {\csname oldend#1\endcsname\endlinenomath}%
}
\newcommand*\patchBothAmsMathEnvironmentsForLineno[1]{%
  \patchAmsMathEnvironmentForLineno{#1}%
  \patchAmsMathEnvironmentForLineno{#1*}%
}
\AtBeginDocument{%
\patchBothAmsMathEnvironmentsForLineno{equation}%
\patchBothAmsMathEnvironmentsForLineno{align}%
\patchBothAmsMathEnvironmentsForLineno{flalign}%
\patchBothAmsMathEnvironmentsForLineno{alignat}%
\patchBothAmsMathEnvironmentsForLineno{gather}%
\patchBothAmsMathEnvironmentsForLineno{multline}%
\patchBothAmsMathEnvironmentsForLineno{eqnarray}%
}

\usepackage[pdftex,
            pdfauthor={\paperauthors},
            pdftitle={\paperasciititle},
            pdfkeywords={\paperkeywords}]{hyperref}
\usepackage{hyperxmp}
\hypersetup{
    pdfcopyright={Copyright (C) \papercopyright},
    pdflicenseurl={\paperlicenceurl}
}

\usepackage[colorinlistoftodos,textsize=scriptsize]{todonotes}

\usepackage[bottom,flushmargin,hang,multiple]{footmisc}

\usepackage[all]{hypcap} 

\input{lhcb-symbols-def} 

\hypersetup{
  colorlinks   = true, 
  urlcolor     = blue, 
  linkcolor    = blue, 
  citecolor    = red   
}

\ifEnableSectionTOCLinks
    \usepackage[explicit]{titlesec} 
    
    \let\oldcontentsline\contentsline
    \renewcommand

    \titleformat{\section}{\normalfont\Large\bf}{\hyperlink{tocsection.\thesection}{{\thesection} \parbox[t]{\dimexpr\textwidth-1pc}{#1}}}{1pc}{}

    \titleformat{\subsection}{\normalfont\bf}{\hyperlink{tocsubsection.\thesubsection}{{\thesubsection} \parbox[t]{\dimexpr\textwidth-1pc}{#1}}}{1pc}{}

    \titleformat{name=\section,numberless}[display]{}{}{0pt}{\normalfont\Huge\bfseries #1}
\fi

\usepackage{cite} 
\usepackage{mciteplus}

%% file: lhcb-symbols-def.tex
\usepackage{xspace}
\usepackage{upgreek}

\def\lhcb   {\mbox{LHCb}\xspace}

\def\besiii {\mbox{BESIII}\xspace}

\def\lhc    {\mbox{LHC}\xspace}

\def\velo   {VELO\xspace}

\def\MagUp {\mbox{\em Mag\kern -0.05em Up}\xspace}

\ifthenelse{\boolean{uprightparticles}}%
{

 \def\Pmu         {\ensuremath{\upmu}\xspace}
 \def\Pnu         {\ensuremath{\upnu}\xspace}
 
 \def\Ppi         {\ensuremath{\uppi}\xspace}

 \def\Ppsi        {\ensuremath{\uppsi}\xspace}

 \def\PDelta      {\ensuremath{\Delta}\xspace}
 \def\PXi         {\ensuremath{\Xi}\xspace}
 \def\PLambda     {\ensuremath{\Lambda}\xspace}
 \def\PSigma      {\ensuremath{\Sigma}\xspace}
 \def\POmega      {\ensuremath{\Omega}\xspace}
 \def\PUpsilon    {\ensuremath{\Upsilon}\xspace}
 \let\oldPi\Pi
 \def\PPi         {\ensuremath{\oldPi}\xspace}

 \def\PB      {\ensuremath{\mathrm{B}}\xspace}
 \def\PD      {\ensuremath{\mathrm{D}}\xspace}
 \def\PJ      {\ensuremath{\mathrm{J}}\xspace}
 \def\PK      {\ensuremath{\mathrm{K}}\xspace}
 \def\Pb      {\ensuremath{\mathrm{b}}\xspace}
 \def\Pc      {\ensuremath{\mathrm{c}}\xspace}
 \def\Pd      {\ensuremath{\mathrm{d}}\xspace}
 \def\Pe      {\ensuremath{\mathrm{e}}\xspace}
 \def\Pp      {\ensuremath{\mathrm{p}}\xspace}

 \def\Ps      {\ensuremath{\mathrm{s}}\xspace}
 
 \def\Pu      {\ensuremath{\mathrm{u}}\xspace}
 \def\thebaroffset{0.0em}
}
{

 \def\Pmu         {\ensuremath{\mu}\xspace}
 \def\Pnu         {\ensuremath{\nu}\xspace}
 
 \def\Ppi         {\ensuremath{\pi}\xspace}

 \def\Ppsi        {\ensuremath{\psi}\xspace}
 
 \mathchardef\PDelta="7101
 \mathchardef\PXi="7104
 \mathchardef\PLambda="7103
 \mathchardef\PSigma="7106
 \mathchardef\POmega="710A
 \mathchardef\PUpsilon="7107
 \mathchardef\PPi="7105
 \def\PB      {\ensuremath{B}\xspace}
 \def\PD      {\ensuremath{D}\xspace}
 \def\PJ      {\ensuremath{J}\xspace}
 \def\PK      {\ensuremath{K}\xspace}
 \def\Pb      {\ensuremath{b}\xspace}
 \def\Pc      {\ensuremath{c}\xspace}
 \def\Pd      {\ensuremath{d}\xspace}
 \def\Pe      {\ensuremath{e}\xspace}
 \def\Pp      {\ensuremath{p}\xspace}

 \def\Ps      {\ensuremath{s}\xspace}
 
 \def\Pu      {\ensuremath{u}\xspace}
 \def\thebaroffset{0.18em}
}
\newcommand{\offsetoverline}[2][\thebaroffset]{\kern #1\overline{\kern -#1 #2}}%

\makeatletter
\ifcase \@ptsize \relax
  \newcommand{\miniscule}{\@setfontsize\miniscule{4}{5}}
\or
  \newcommand{\miniscule}{\@setfontsize\miniscule{5}{6}}
\or
  \newcommand{\miniscule}{\@setfontsize\miniscule{5}{6}}
\fi
\makeatother

\DeclareRobustCommand{\optbar}[1]{\shortstack{{\miniscule (\rule[.5ex]{1.25em}{.18mm})}
  \\ [-.7ex] $#1$}}

\def\electron   {{\ensuremath{\Pe}}\xspace}

\def\muon       {{\ensuremath{\Pmu}}\xspace}
\def\mup        {{\ensuremath{\Pmu^+}}\xspace}
\def\mun        {{\ensuremath{\Pmu^-}}\xspace} 

\def\neub       {{\ensuremath{\overline{\Pnu}}}\xspace}

\def\neueb      {{\ensuremath{\neub_e}}\xspace}

\def\neumb      {{\ensuremath{\neub_\mu}}\xspace}

\def\neutb      {{\ensuremath{\neub_\tau}}\xspace}

\def\uquark    {{\ensuremath{\Pu}}\xspace}

\def\dquark    {{\ensuremath{\Pd}}\xspace}

\def\squark    {{\ensuremath{\Ps}}\xspace}

\def\cquark    {{\ensuremath{\Pc}}\xspace}

\def\bquark    {{\ensuremath{\Pb}}\xspace}

\def\pion   {{\ensuremath{\Ppi}}\xspace}

\def\pip    {{\ensuremath{\pion^+}}\xspace}
\def\pim    {{\ensuremath{\pion^-}}\xspace}

\def\kaon    {{\ensuremath{\PK}}\xspace}

\def\KorKbar {\kern \thebaroffset\optbar{\kern -\thebaroffset \PK}{}\xspace}

\def\KS      {{\ensuremath{\kaon^0_{\mathrm{S}}}}\xspace}

\def\D       {{\ensuremath{\PD}}\xspace}

\def\DorDbar {\kern \thebaroffset\optbar{\kern -\thebaroffset \PD}\xspace}

\def\Dp      {{\ensuremath{\D^+}}\xspace}
\def\Dm      {{\ensuremath{\D^-}}\xspace}

\def\DpDm    {\ensuremath{\Dp {\kern -0.16em \Dm}}\xspace}

\def\B       {{\ensuremath{\PB}}\xspace}

\def\BorBbar {\kern \thebaroffset\optbar{\kern -\thebaroffset \PB}\xspace}

\def\Bd      {{\ensuremath{\B^0}}\xspace}

\def\BdorBdbar {\kern \thebaroffset\optbar{\kern -\thebaroffset \Bd}\xspace}

\def\Bs      {{\ensuremath{\B^0_\squark}}\xspace}

\def\BsorBsbar {\kern \thebaroffset\optbar{\kern -\thebaroffset \Bs}\xspace}

\def\jpsi     {{\ensuremath{{\PJ\mskip -3mu/\mskip -2mu\Ppsi}}}\xspace}

\def\Y#1S{\ensuremath{\PUpsilon{(#1S)}}\xspace}

\def\proton      {{\ensuremath{\Pp}}\xspace}

\def\Lz          {{\ensuremath{\PLambda}}\xspace}
\def\Lbar        {{\ensuremath{\offsetoverline{\PLambda}}}\xspace}
\def\LorLbar     {\kern \thebaroffset\optbar{\kern -\thebaroffset \PLambda}\xspace}

\def\BF         {{\ensuremath{\mathcal{B}}}\xspace}

\newcommand{\decay}[2]{\mbox{\ensuremath{#1\!\to #2}}\xspace}

\def\to                 {\ensuremath{\rightarrow}\xspace}

\def\Vud  {{\ensuremath{V_{\uquark\dquark}^{\phantom{\ast}}}}\xspace}

\def\Vus  {{\ensuremath{V_{\uquark\squark}^{\phantom{\ast}}}}\xspace}

\def\Vub  {{\ensuremath{V_{\uquark\bquark}^{\phantom{\ast}}}}\xspace}

\def\AT#1     {\ensuremath{A_{\mathrm{T}}^{#1}}\xspace}           

\def\C#1      {\ensuremath{\mathcal{C}_{#1}}\xspace}                       
\def\Cp#1     {\ensuremath{\mathcal{C}_{#1}^{'}}\xspace}                    
\def\Ceff#1   {\ensuremath{\mathcal{C}_{#1}^{\mathrm{(eff)}}}\xspace}        
\def\Cpeff#1  {\ensuremath{\mathcal{C}_{#1}^{'\mathrm{(eff)}}}\xspace}       
\def\Ope#1    {\ensuremath{\mathcal{O}_{#1}}\xspace}                       
\def\Opep#1   {\ensuremath{\mathcal{O}_{#1}^{'}}\xspace}                    

\newcommand{\nospaceunit}[1]{\ensuremath{\text{#1}}}
\newcommand{\aunit}[1]{\ensuremath{\text{\,#1}}}

\newcommand{\tev}{\aunit{Te\kern -0.1em V}\xspace}
\newcommand{\gev}{\aunit{Ge\kern -0.1em V}\xspace}
\newcommand{\mev}{\aunit{Me\kern -0.1em V}\xspace}
\newcommand{\kev}{\aunit{ke\kern -0.1em V}\xspace}
\newcommand{\ev}{\aunit{e\kern -0.1em V}\xspace}

\newcommand{\mevc}{\ensuremath{\aunit{Me\kern -0.1em V\!/}c}\xspace}
\newcommand{\gevc}{\ensuremath{\aunit{Ge\kern -0.1em V\!/}c}\xspace}
\newcommand{\mevcc}{\ensuremath{\aunit{Me\kern -0.1em V\!/}c^2}\xspace}
\newcommand{\gevcc}{\ensuremath{\aunit{Ge\kern -0.1em V\!/}c^2}\xspace}

\def\mum  {\ensuremath{\,\upmu\nospaceunit{m}}\xspace}

\def\fb   {\ensuremath{\aunit{fb}}\xspace}
\def\invfb   {\ensuremath{\fb^{-1}}\xspace}

\def\ps   {\ensuremath{\aunit{ps}}\xspace}

\newcommand{\stat}{\aunit{(stat)}\xspace}
\newcommand{\syst}{\aunit{(syst)}\xspace}

\def\gsim{{~\raise.15em\hbox{$>$}\kern-.85em
          \lower.35em\hbox{$\sim$}~}\xspace}
\def\lsim{{~\raise.15em\hbox{$<$}\kern-.85em
          \lower.35em\hbox{$\sim$}~}\xspace}

\def\pt         {\ensuremath{p_{\mathrm{T}}}\xspace}

\def\ptot       {\ensuremath{p}\xspace}

\def\evtgen     {\mbox{\textsc{EvtGen}}\xspace}

\def\geant      {\mbox{\textsc{Geant4}}\xspace}

\def\photos     {\mbox{\textsc{Photos}}\xspace}

\def\pythia     {\mbox{\textsc{Pythia}}\xspace}

\def\tell1  {TELL1\xspace}
\def\ukl1   {UKL1\xspace}

\newcommand{\lhcborcid}[1]{\href{https://orcid.org/#1}{\hspace*{0.1em}\raisebox{-0.45ex}{\includegraphics[width=1em]{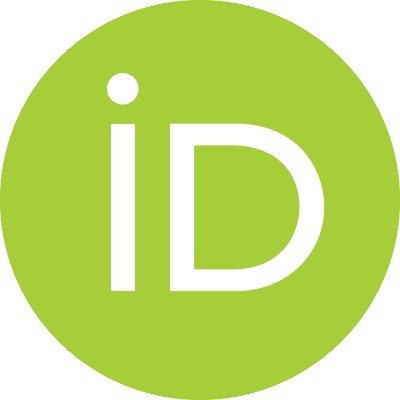}}}}


%% file: title-LHCb-PAPER.tex

\begin{titlepage}
\pagenumbering{roman}

\vspace*{-1.5cm}
\centerline{\large EUROPEAN ORGANIZATION FOR NUCLEAR RESEARCH (CERN)}
\vspace*{1.5cm}
\noindent
\begin{tabular*}{\linewidth}{lc@{\extracolsep{\fill}}r@{\extracolsep{0pt}}}
\ifthenelse{\boolean{pdflatex}}
{\vspace*{-1.5cm}\mbox{\!\!\!\includegraphics[width=.14\textwidth]{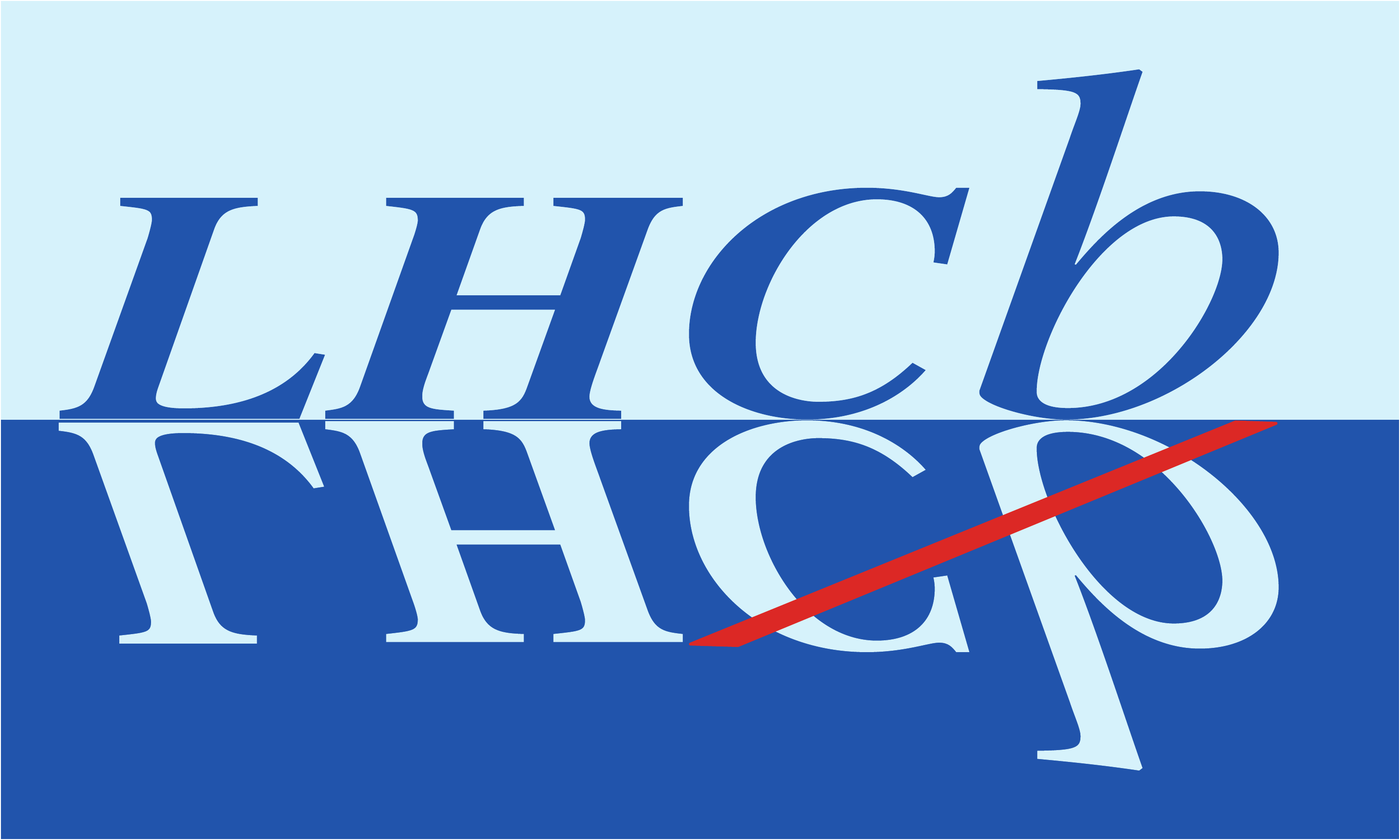}} & &}%
{\vspace*{-1.2cm}\mbox{\!\!\!\includegraphics[width=.12\textwidth]{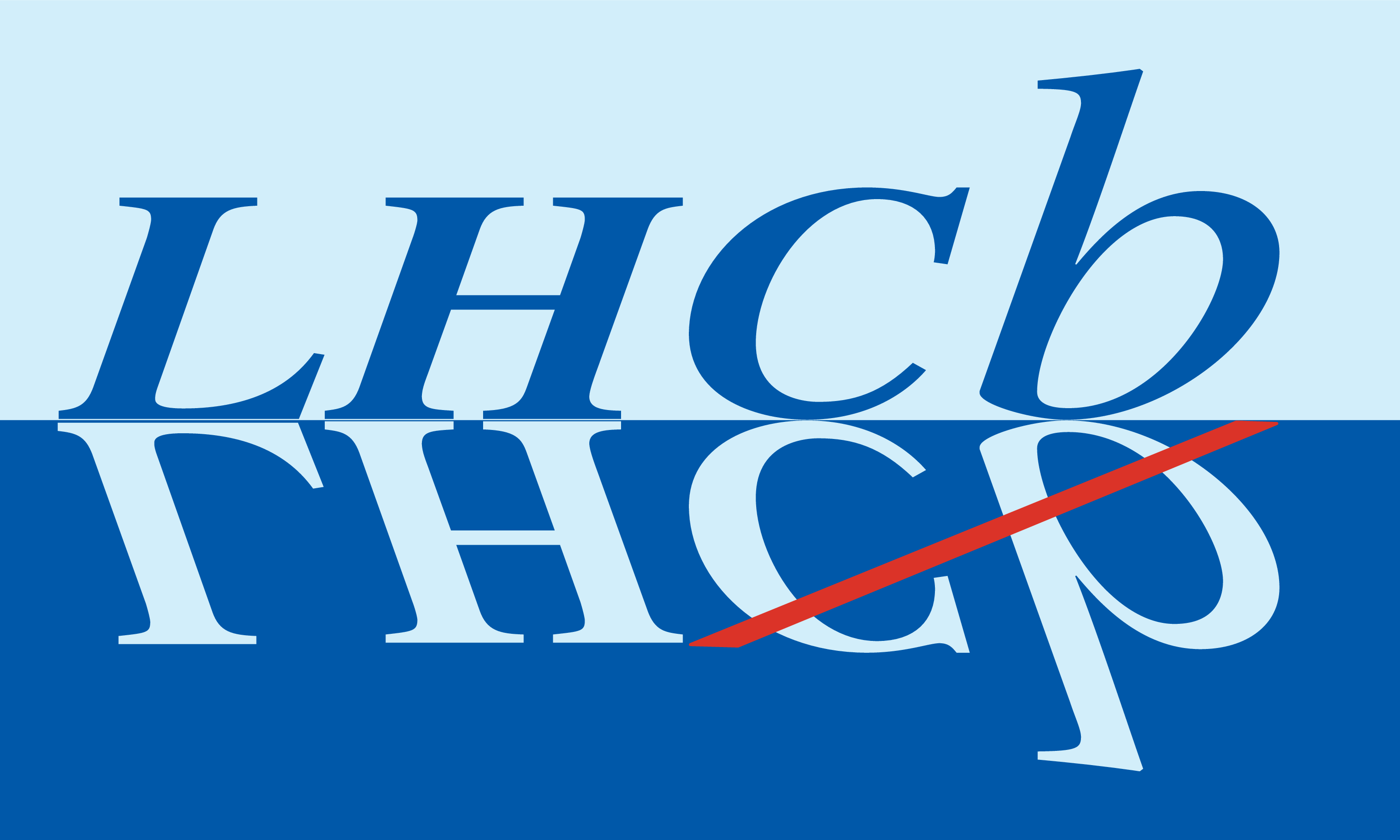}} & &}%
\\
 & & CERN-EP-2025-223 \\  
 & & LHCb-PAPER-2025-030 \\  
 & & April 20, 2026 \\ 
 & & \\
\end{tabular*}

\vspace*{4.0cm}

{\normalfont\bfseries\boldmath\huge
\begin{center}
  \papertitle 
\end{center}
}

\vspace*{1.6cm}

\begin{center}
\paperauthors\footnote{Authors are listed at the end of this paper.}
\end{center}

\vspace{\fill}

\begin{abstract}
  \noindent
    A measurement of the branching fraction for the decay \decay{\Lz}{\proton \mun \neumb}  is presented using \textit{pp} collision data collected by the \lhcb experiment at a centre-of-mass energy of 13\tev. The analysis is based on data recorded between 2016 and 2018, corresponding to an integrated luminosity of $5.4\invfb$. The result is obtained using \decay{\Lz}{\proton \pim} decays as a normalisation channel. The measured branching fraction is \mbox{$\mathcal{B}(\decay{\Lz}{\proton \mun \neumb} )= (1.462 \pm 0.016 \pm 0.100 \pm 0.011 ) \times 10^{-4}$}, where the uncertainties are statistical, systematic, and due to the limited knowledge of the normalisation mode branching fraction, respectively. This result improves the precision of the branching fraction measurement by a factor of two compared to the previous best measurement and sets a more stringent bound on lepton flavour universality in \decay{\squark}{\uquark} quark transitions. It is consistent with previous measurements, and the extracted lepton flavour universality test observable, $R^{\mu e} = \frac{\Gamma(\Lambda \to p \mu^- \bar{\nu}_\mu)}{\Gamma(\Lambda \to p e^- \bar{\nu}_e)} =  0.175 \pm 0.012$,  agrees with the Standard Model prediction.
  
\end{abstract}

\vspace*{1.6cm}

\begin{center}
  Published in Journal of High Energy Physics 04 (2026) 096
\end{center}

\vspace{\fill}

{\footnotesize 
\centerline{\copyright~\papercopyright. \href{\paperlicenceurl}{\paperlicence}.}}
\vspace*{2mm}

\end{titlepage}


\newpage
\setcounter{page}{2}
\mbox{~}
%
%
%
%

%% file: body.tex
\section{Introduction}
\label{sec:Introduction}

Precise measurements of \decay{\bquark}{\cquark} transitions have provided hints of lepton flavour universality (LFU) violation, which could indicate physics beyond the Standard Model (SM)~\cite{Nature-546,HFLAV21}. While the most significant deviations
have so far been observed in $\decay{\bquark}{\cquark \tau^- \neutb}$ decays, there has
also been sustained interest in testing LFU using ratios of light leptons in
tree-level charged-current processes, where theoretical uncertainties are
particularly well controlled~\cite{Bryman:2022}. No such effects have been observed so far in other charged-current decays of down-type quarks. This paper reports the most precise test to date of LFU in charged-current \decay{\squark}{\uquark} hyperon decays. The \lhcb experiment has demonstrated excellent capabilities in the study of strange hadron decays, particularly in searches for rare processes. The collaboration has reported the most restrictive upper limits on the branching fractions of the \decay{\KS}{\mup \mun} and \decay{\KS}{\mup \mun \mup \mun} decays, and has performed the most precise measurement to date of the \decay{\PSigma^+}{\proton \mup \mun} branching fraction~\cite{LHCb-PAPER-2019-038,LHCb-PAPER-2022-035,LHCb-PAPER-2025-002}.

From a theoretical standpoint, semileptonic hyperon decays (SHDs) have been shown to be sensitive probes of specific beyond-the-SM dynamics that could break LFU~\cite{Chang:2014iba,TheBetaDecaysOfHyperons}. The differential decay rates of these processes are governed by the parameter $\delta \equiv \frac{m_1 - m_2}{m_1} = \frac{\Delta}{m_1} $, where $m_{1,2}$ are the masses of the initial ($B_1$) and final ($B_2$) state baryons, and $\Delta \equiv m_1 - m_2$ denotes their mass difference. The $\delta$ parameter quantifies the degree of SU(3) flavour symmetry breaking in SHDs. Compared to electron final states, decays with muons in the final state are particularly sensitive to scalar and tensor contributions, which are suppressed by a factor of $m_\ell / \sqrt{q^2}$, where $m_\ell$ is the mass of the charged lepton and $q^2$ is the squared momentum transferred to the leptonic system. Precision measurements in this sector therefore complement direct searches for new phenomena at the \lhc~\cite{Chang:2014iba}.

The approximate SU(3) flavour symmetry present in hyperons constrains the decay phase space and permits an expansion of the decay rate in terms of the SU(3) symmetry-breaking parameter $\delta$, which is approximately 0.159 for the SHD of a $\Lz$ particle, when the final-state baryon is a proton~\cite{Chang:2014iba, TheBetaDecaysOfHyperons}. For convenience, the decay rate is first expressed for the electronic mode, where the
lepton mass can be neglected ($m_e \ll \Delta$), and the corresponding muonic decay rate is subsequently
obtained through the decay rates ratio $R^{\mu e}$, which explicitly accounts for the finite muon
mass. The integrated $B_1\to B_2 e^- \bar{\nu}_e$ decay rate --- assuming real form factors and expanding in $\delta$ up to next-to-leading order (NLO) --- is given by~\cite{SHD}

\begin{equation}
    \label{eq:DecayRateElectron}
    \Gamma_{\text{NLO}}(B_1\to B_2 e^- \bar{\nu}_e) \simeq \frac{G_F^2 | \Vus f_1(0) |^2 \Delta^5}{60 \pi^3} \left[ \left( 1 - \frac{3}{2} \delta \right) \left( 1 + 3 \frac{g_1(0)^2}{f_1(0)^2} \right) - 4 \delta \frac{g_1(0)g_2(0)}{f_1(0)^2} \right]. 
\end{equation}
In this expression, $f_1(0) $ and $g_1(0)$ represent the vector and axial-vector charges of the hyperon, corresponding to the respective vector and axial-vector form factors evaluated at zero momentum transfer ($q^2 = 0$), while $g_2(0)$ denotes the weak-electric charge. Here, $G_F$ denotes the Fermi constant, and \Vus is the CKM matrix element. Notably, the decay rate exhibits only a weak dependence on the form factors beyond their values at $q^2 = 0$ and their $ q^2$ dependence is expected to play a subleading role at this level of approximation. Furthermore, the contribution from the weak-electric form factor $g_2(0)$ can be neglected, as it is suppressed by symmetry-breaking effects of order $\delta$. Consequently, the SM prediction for the total decay rate in the electronic mode depends only on the hyperon vector and axial charges, $f_1(0)$ and $g_1(0)$. The neglected terms of order $\mathcal{O}(\delta^2)$ in the SU(3)-breaking expansion are estimated to affect the predicted decay rate at the level of 1 to 5\%~\cite{Chang:2014iba}.

The LFU test observable ($R^{\mu e}$), defined as the ratio of the decay rates of the muon and electron modes, is sensitive to nonstandard scalar and tensor contributions~\cite{Chang:2014iba}. Moreover, the dependence on the form factors is anticipated to simplify when considering the ratio of branching fractions. By operating at NLO one obtains

\begin{equation}
\label{eq:SMratio}
    \begin{split}
    R_{\text{NLO}}^{\mu e} = \sqrt{1-\frac{m_{\mu}^2}{\Delta^2}} \Bigg( 1 - \frac{9}{2} \frac{m_{\mu}^2}{\Delta^2} - 4 \frac{m_{\mu}^4}{\Delta^4}  \Bigg) + \frac{15}{2} \frac{m_{\mu}^4}{\Delta^4} \mathrm{arctanh} \Bigg( \sqrt{1-\frac{m_{\mu}^2}{\Delta^2}}\Bigg) \, .
    \end{split}
\end{equation}
The LFU test observable defined in Eq.~\ref{eq:SMratio} is predicted by theory to be  \mbox{$R_{\text{NLO}}^{\mu e} = 0.153 \pm 0.008$} for the $\Lz \to \proton$ case~\cite{Chang:2014iba}.\footnote{\label{note0}Note that the decay rate for the muon mode can be obtained by multiplying that of the electron mode by $R_{\text{NLO}}^{\mu e}$, as defined in Eq.~\ref{eq:SMratio}.} This prediction is theoretically clean, as it does not depend on form-factor inputs at this perturbative order, allowing for a straightforward comparison with experimental measurements. A precise measurement of $R^{\mu e}$ implies a constraint on the Wilson coefficients, since new physics scalar and tensor operators would contribute to the ratio. Long-distance QED effects are not explicitly included in these predictions and are not expected to affect the results at the current level of experimental precision \cite{Chang:2014iba}.

A recent lattice QCD calculation provides the most precise determination to date of the $\Lz \to \proton \, \ell^- \bar{\nu}_\ell$ transition form factors, including previously neglected second-class contributions, yielding a value of the muon-to-electron decay rate ratio of \mbox{$R_{\text{LQCD}}^{\mu e} = 0.1735 \pm 0.0098$}~\cite{LQCD}. In addition to the form-factor determination, the same study establishes a precise relationship between the partial decay rates of the $\Lz \to \proton \, \ell^- \bar{\nu}_\ell$ decays and the CKM matrix element $|V_{us}|$. Specifically, by integrating the differential decay rates using only lattice QCD inputs—with $|V_{us}|$ factored out—predictions are provided for the ratios $\Gamma_{p\ell \bar{\nu}_\ell} / |V_{us}|^2$ for both the electron and muon modes. Two sets of predictions~\cite{LQCD} are reported: one using baryon masses determined within the same lattice QCD calculation, and another using experimentally measured baryon masses. The former adopts a more conservative uncertainty, intended as the main quoted value, to account for potential uncontrolled systematics in the baryon spectrum and yields
\begin{align}
    \frac{\Gamma_{pe\bar{\nu}_e}}{|V_{us}|^2} &= (5.83 \pm 0.37) \times 10^7 \, \mathrm{s}^{-1}, \\
    \frac{\Gamma_{p\mu\bar{\nu}_\mu}}{|V_{us}|^2} &= (1.01 \pm 0.12) \times 10^7 \, \mathrm{s}^{-1}, \label{eq:LQCD-lattice}
\end{align}
while the predictions obtained using experimental baryon masses, providing smaller uncertainties but without a full treatment of the systematic effects, are
\begin{align}
    \frac{\Gamma_{pe\bar{\nu}_e}}{|V_{us}|^2} &= (5.546 \pm 0.051) \times 10^7 \, \mathrm{s}^{-1}, \\
    \frac{\Gamma_{p\mu\bar{\nu}_\mu}}{|V_{us}|^2} &= (0.9228 \pm 0.0082) \times 10^7 \, \mathrm{s}^{-1}, \label{eq:LQCD-lattice2}\\
    R_{\text{LQCD}}^{\mu e} &= 0.16638 \pm 0.00020. \label{eq:LQCD-LFU}
\end{align}

A precise determination of the CKM matrix element $|\Vus|$ is essential for testing the unitarity of the first row of the CKM matrix, a fundamental property of the SM. The most accurate measurements of $|\Vus|$ come from kaon decays: leptonic ($K_{\mu2}$) and semileptonic ($K_{\ell3}$) channels. However, these two approaches yield results that differ at the $3\sigma$ level~\cite{PDG2024,Hidden}. When the $|\Vus|$ measurements are combined with the precisely measured value of $|\Vud|$, given that the contribution from \Vub element is almost entirely negligible
(approximately $1.3 \times 10^{-5}$), the unitarity condition $|\Vud|^2 + |\Vus|^2 + |\Vub|^2 = 1$ shows a deviation at the $2\sigma$ level \cite{PDG2024}. This so-called ``Cabibbo angle anomaly'' has motivated renewed interest in precision measurements of $|\Vus|$.

Prospects for studying strange-hadron decays at \lhcb, highlighting the \mbox{\decay{\Lz}{\proton \mun \neumb}} channel\footnote{\label{note}Charge conjugation is implied throughout this paper.} as particularly promising due to its high reconstruction efficiency, are motivated in Ref.~\cite{JHEP-2019-48}. In 2021, the \besiii collaboration reported the first measurement of the absolute branching fraction for this decay, obtaining $\BF(\decay{\Lz}{\proton \mun \neumb}) = (1.48 \pm 0.21) \times 10^{-4}$~\cite{BESIII:2021ynj}, which remains the most precise result from a single measurement to date. The current world average is $(1.51 \pm 0.19) \times 10^{-4}$~\cite{PDG2024}. This paper presents a new measurement of the \decay{\Lz}{\proton \mun \neumb} branching fraction with improved precision, and the first study of lepton flavour universality in hyperon decays at \lhcb.

\section{Detector and simulation}

The data were collected in proton-proton collisions by the \lhcb experiment at a centre-of-mass energy of 13\,\tev between 2016 and 2018, corresponding to an integrated luminosity of 5.4\,\invfb. The \lhcb detector~\cite{LHCb-DP-2008-001,LHCb-DP-2014-002} is a single-arm forward spectrometer covering the \mbox{pseudorapidity} range $2<\eta <5$,
designed for the study of particles containing \bquark or \cquark
quarks. The detector used to collect the data analysed in this paper  includes a high-precision tracking system
consisting of a silicon-strip vertex detector surrounding the $pp$
interaction region, the \velo, a large-area silicon-strip detector located
upstream of a dipole magnet with a bending power of about
$4{\mathrm{\,T\,m}}$, and three stations of silicon-strip detectors and straw
drift tubes placed downstream of the magnet.

The tracking system provides a measurement of the momentum, \ptot, of charged particles with
a relative uncertainty that varies from 0.5\% at low momentum to 1.0\% at 200\gevc.
The minimum distance of a track to a primary $pp$ collision vertex (PV), the impact parameter (IP), 
is measured with a resolution of $(15+29/\pt)\mum$,
where \pt is the component of the momentum transverse to the beam, in\,\gevc.
Different types of charged hadrons are distinguished using information
from two ring-imaging Cherenkov detectors. 
Photons, electrons and hadrons are identified by a calorimeter system consisting of
scintillating-pad and preshower detectors, an electromagnetic
and a hadronic calorimeter. Muons are identified by a system composed of alternating layers of iron and multiwire proportional chambers. The online event selection is performed by a trigger, which consists of a hardware stage, based on information from the calorimeter and muon systems, followed by a software stage, which applies a full event reconstruction. Triggered data further undergo a centralised, offline processing step, referred to here as the preselection,
to deliver physics-analysis-ready data across the entire \lhcb physics programme~\cite{Stripping}.

Simulation is required to model the effects of the detector acceptance and the imposed selection requirements. In the simulation, $pp$ collisions are generated using \pythia~\cite{Sjostrand:2007gs,*Sjostrand:2006za} with a specific \lhcb configuration~\cite{LHCb-PROC-2010-056}.
Decays of unstable particles are described by \evtgen~\cite{Lange:2001uf}, in which final-state radiation is generated using \photos~\cite{davidson2015photos}. The interaction of the generated particles with the detector, and its response, are implemented using the \geant toolkit~\cite{Allison:2006ve, *Agostinelli:2002hh} as described in  Ref.~\cite{LHCb-PROC-2011-006}.

Three main simulated samples are used in this analysis: the signal decay, \decay{\Lz}{\proton \pim} decays, and a minimum-bias sample generated to emulate collision events selected by a dedicated random trigger in data. The signal decays are simulated using a custom \evtgen model, that implements the theoretical differential decay rate of SHDs~\cite{Kadeer_2008}, in order to reproduce the correct kinematic distributions. These samples are weighted to account for observed differences in the $\Lz$ transverse momentum and pseudorapidity distributions relative to data (see Sec.~\ref{subsec:corrections}).


\section{Analysis overview}

The measurement of the branching ratio of the decay \decay{\Lz}{\proton \mun \neumb} is performed relative to the \decay{\Lz}{\proton \pim}  decay, whose branching ratio has been measured as \mbox{$\BF(\decay{\Lz}{\proton \pim})=0.641 \pm 0.005$}~\cite{PDG2024}. Two sets of preselection criteria are applied: one for the normalisation channel and the other for the signal channel. The two preselections are designed to be as similar as possible, such that systematic uncertainties largely cancel in the ratio. The only differences arise from particle-identification requirements and the mass hypothesis used in the reconstruction. A detailed description of the preselection criteria is given in Sec.~\ref{sec:CandidatesSelection}, which also describes the selection requirements used to further isolate the signal from physical backgrounds arising from \decay{\Lz}{\proton \pim} and $\KS \to \pip \pim$ decays.

The \decay{\Lz}{\proton \mun \neumb} branching ratio can be expressed as

\begin{equation}
    \label{eq:number1}
    \BF(\decay{\Lz}{\proton \mun \neumb}) = \BF(\decay{\Lz}{\proton \pim})\frac{\epsilon_{\Lz \to p\pi}}{\epsilon_{\Lz \to p\muon\neumb}}\frac{N_{\Lz \to p\muon\neumb}}{N_{\Lz \to p\pi}} = \alpha N_{\Lz \to p\muon\neumb}\ ,
\end{equation}
where $\epsilon_{\Lz \to p\pi}$ and $\epsilon_{\Lz \to p\muon\neumb}$ are the selection efficiencies for the normalisation and signal channels, respectively, and $N_{\Lz \to p\pi}$ and $N_{\Lz \to p\muon\neumb}$ are the corresponding yields. All values except for $N_{\Lz \to p\muon\neumb}$ are included in the normalisation constant $\alpha$.

In the offline selection, trigger signals are associated with reconstructed particles. Selection requirements can therefore be made on the trigger selection itself and on whether the decision was due to the signal candidate, other particles produced in the $pp$ collision, or a combination of both. Thanks to the high $\Lz$ production rate at the LHC---approximately one $\Lz$ per primary interaction~\cite{JHEP-2019-48}--- this analysis can be performed selecting only candidates that are triggered independent of signal (TIS). Although this requirement reduces the overall trigger efficiency, it ensures consistency between the signal and normalisation modes, thereby reducing the impact of systematic uncertainties.



\section{Variable definitions}
\label{sec:Variables}

The minimum-bias simulation sample is used to study the background composition after applying the  signal candidate preselection described in Sec.~\ref{sec:CandidatesSelection}. It is found that approximately 60\% of selected \decay{\Lz}{\proton \mun \neumb} candidates originate from either \decay{\Lz}{\proton \pim} or 
\decay{\Lz}{\proton \pim(\decay{}{\mun \neumb})} decays, where the pion decays in flight. Given the similarity between the background and signal topologies, a dedicated strategy is optimised to enhance their separation. This involves exploiting the distinct kinematic features of each decay mode to construct discriminating variables for the candidates passing the signal preselection.

\begin{figure}[tb]
  \begin{center}
    \includegraphics[width=0.495\linewidth]{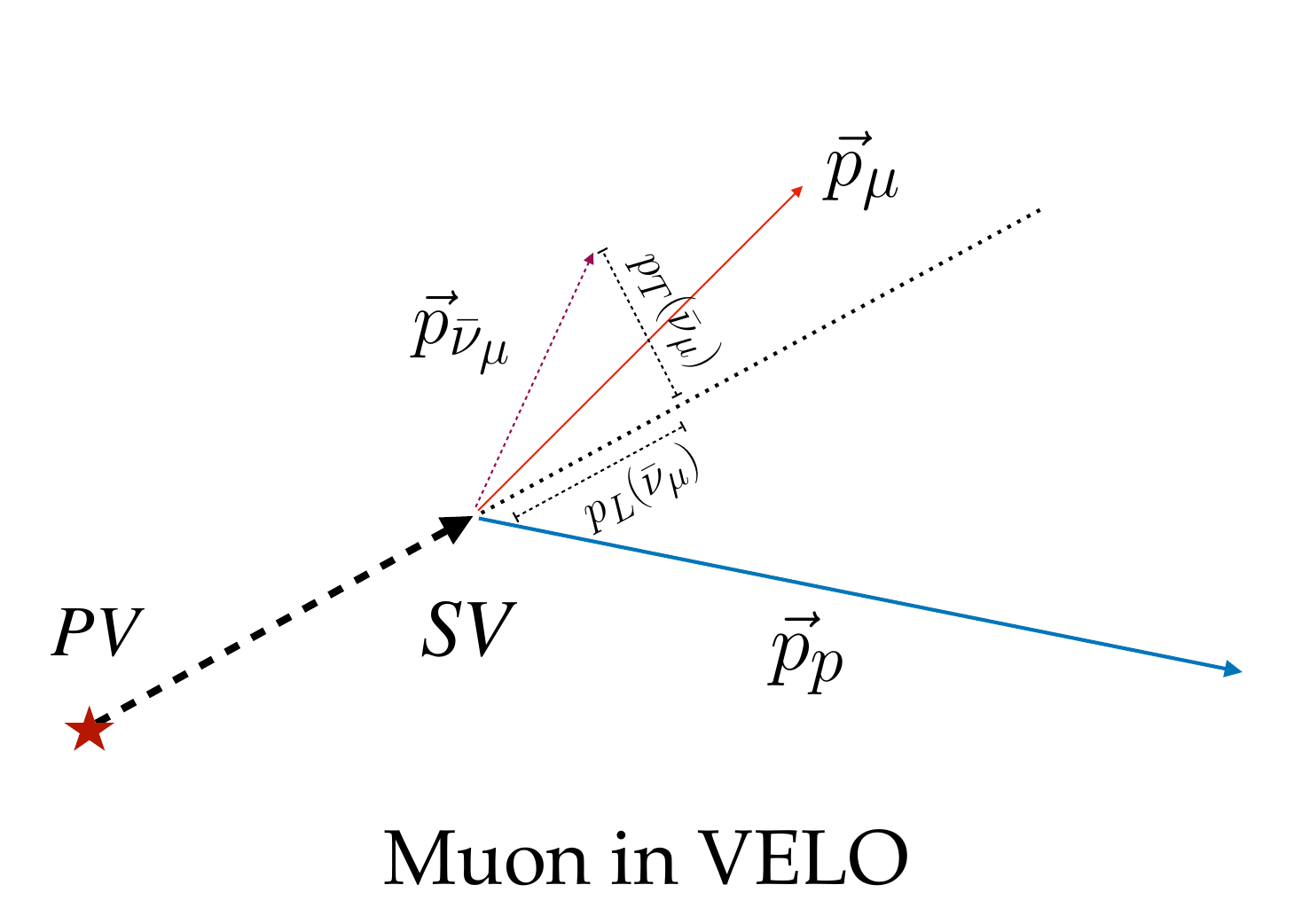}
    \includegraphics[width=0.495\linewidth]{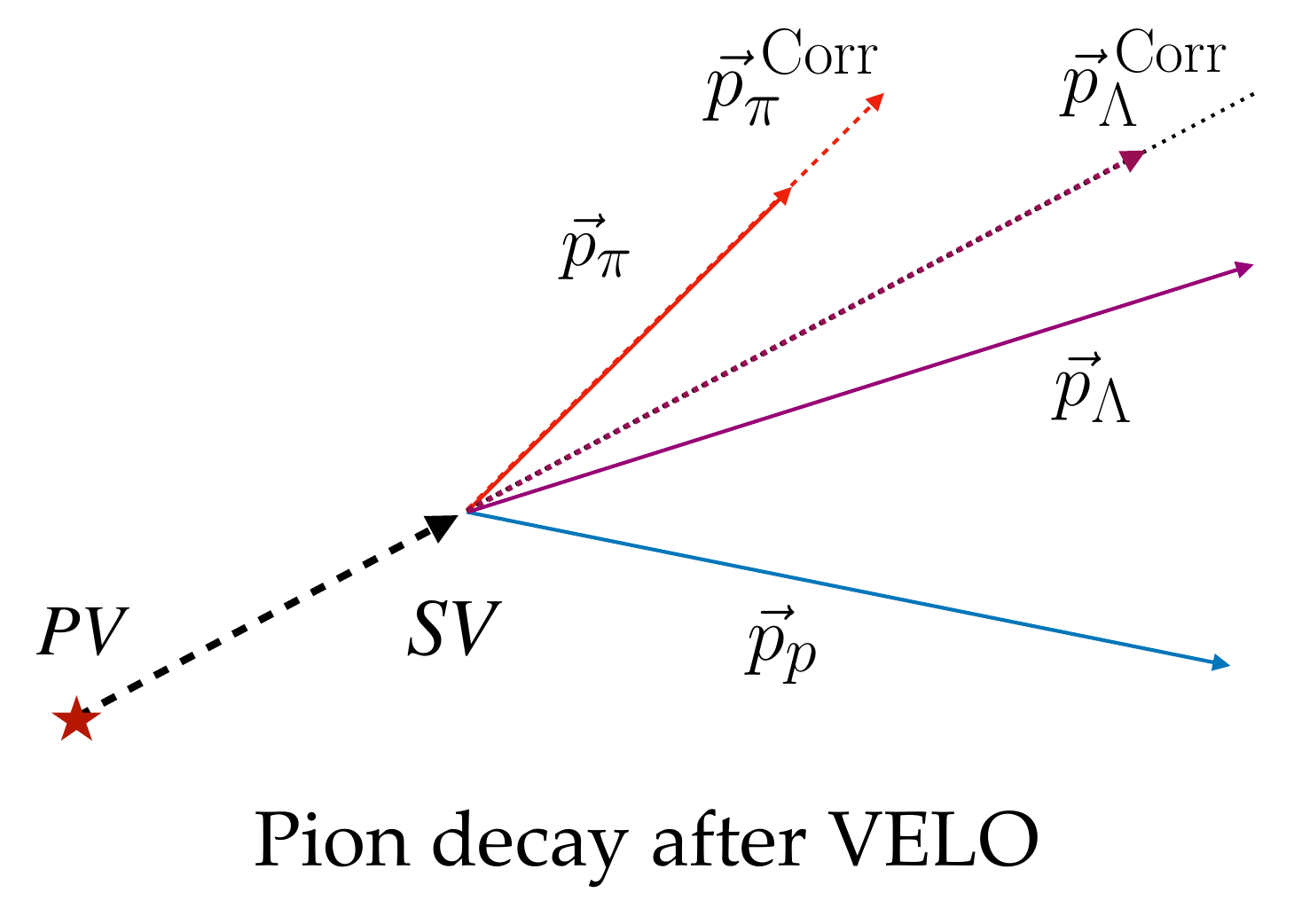}
  \end{center}
  \caption{
    Topology of the muon and pion categories at \velo level, with an illustration of the kinematic strategies used to recover missing information due to the presence of a neutrino. In the  case shown on the left, the neutrino momentum is determined using the \Lz flight direction and imposing the \Lz mass. In the case shown on the right, the pion momentum is corrected so that, when added to the proton momentum, the total points back to the PV.}
  \label{fig:Variables1}
\end{figure}

Since the decay \decay{\Lz}{\proton \pim(\decay{}{\mun \neumb})} constitutes the main source of background for this measurement, the first step is to classify  such decays---misidentified as signal candidates---into two categories depending on the pion decay point. In the first case, the pion decays within the \velo volume, and the resulting muon leaves hits in that subdetector. In the second case, the pion traverses the entire \velo before decaying, and the hits in the subdetector originate exclusively from the pion. The topology of the two categories is shown in Fig.~\ref{fig:Variables1}.

\subsection{Longitudinal neutrino momentum} 

In the first case, which occurs in approximately 35\% of the selected \decay{\Lz}{\proton \pim(\decay{}{\mun \neumb})} decays, the pion decays into a muon and a neutrino within the \velo volume.
In both these cases and in signal decays, the neutrino transverse momentum, $\pt(\neumb)$, can be inferred from the momentum components of the proton and the muon. This is done using the $\Lz$ flight direction, defined as $\vec{f} = \vec{SV} - \vec{PV}$, where $\vec{PV}$ and $\vec{SV}$ denote the positions of the primary and secondary vertices, respectively. The Gram--Schmidt procedure~\cite{Schmidt1907} is then applied to construct an orthonormal basis: $\hat{x}_u = (-\hat{f}_y, +\hat{f}_x, 0)$ and $\hat{y}_u = \hat{f} \times \hat{x}_u$. This orthogonal basis allows for the reconstruction of the missing transverse momentum component perpendicular to the $\Lz$ flight direction, using the following decomposition for the visible momentum:
\begin{equation}
    \label{eq:Variables0}
    \vec{p\,}^{\prime}_{p\muon} = ( \vec{p}_{p\muon} \cdot \hat{x}_u,\, \vec{p}_{p\muon} \cdot \hat{y}_u,\, \vec{p}_{p\muon} \cdot \hat{f}) 
    = (p^{\prime}_{p\muon,x},\, p^{\prime}_{p\muon,y},\, p^{\prime}_{p\muon,z}) ,
\end{equation}
where
\begin{equation}
    \pt(\neumb) = \sqrt{p_{p\muon,x}^{\prime2} + p_{p\muon,y}^{\prime2}}.
\end{equation}

After computing $\pt(\neumb)$ from the projections of the proton and muon momenta onto the plane transverse to the $\Lz$ flight direction, the longitudinal component $p_{\rm L}(\neumb)$ can be determined by requiring that the invariant mass of the decay products equals the known $\Lz$ mass

\begin{equation}
    \label{eq:Variables2}
    m_{\Lz}^2 = (E_{p\muon} + E_\neumb)^2 - | \vec{p}_{p\muon} + \vec{p}_{\neumb} | ^2 = m_{p\muon}^2 + 2 (E_{p\muon}  E_\neumb - \vec{p}_{p\muon} \cdot \vec{p}_{\neumb}).
\end{equation}
By incorporating
\begin{equation}
    \label{eq:Variables5}
    \vec{p}_{p\muon} \cdot \vec{p}_{\neumb} = - \pt(\neumb)^2 +  p_{p\muon,z}^{\prime}  p_{\rm L}(\neumb) ,
\end{equation}
the longitudinal neutrino momentum is given by

\begin{equation}
    \label{eq:Variables6}
    p_{\rm L}(\neumb) = \frac{E_{p\muon} \sqrt{A^2 - m_\Lz^2 \, \pt(\neumb)^2} - A \, p_{p\muon,z}^{\prime} + p_{p\muon,z}^{\prime} \, \pt(\neumb)^2}{(p_{p\muon,z}^{\prime})^2 - E_{p\muon}^2},
\end{equation}
where
\begin{equation}
    \label{eq:Variables7}
    A=\frac{m_{\Lz}^2-m_{p\muon}^2}{2}.
\end{equation}

 The longitudinal neutrino momentum is a key variable for separation of signal candidates from other decay channels. Signal decays yield a positive value inside the square root of Eq.~\ref{eq:Variables6}, whereas this is typically not the case for combinatorial background, which arises from random combinations of tracks that mimic the signal topology but do not originate from a common decay. In contrast, \decay{\Lz}{\proton \pim}  decays without a decaying pion are expected to give $p_{\rm L}(\neumb)=0$, with some variation due to the finite resolution of the momentum measurement. The requirement of a positive argument inside the square root in Eq.~\ref{eq:Variables6} (see Sec. \ref{sec:CandidatesSelection}, Fig. \ref{fig:MppiCorr}) corresponds to the condition $\pt(\neumb) < (m_{\Lz}^2 - m_{p\mu}^2) / 2 m_{\Lz}$. Signal decays satisfy this condition by construction, whilst combinatorial background typically does not. Background decays of \decay{\Lz}{\proton\pim} yield values close to the boundary, althought events in which the pion decays in flight are more likely to satisfy the condition.

\subsection{Corrected mass}

In the second case, when the pion decays to a muon after the VELO, leaving enough hits in this subdetector to reconstruct a track segment, its momentum direction is known, but the measured momentum modulus does not correspond to the total pion momentum. In such cases, the reconstructed $\Lz$ momentum vector does not correctly point back to the primary vertex. 
The correct $\vec{p}_\pi^{\text{  Corr}}$ can be computed requiring the $\Lz$ momentum to point back to the PV. 

For \decay{\Lz}{\proton \pim(\decay{}{\mun \neumb})} decays, the corrected $\proton\pi$ invariant mass, $ m_{\text{Corr}}(p\pi)$, computed using the obtained value of $\vec{p}_\pi^{\text{  Corr}}$, peaks at the known $\Lz$ mass~\cite{PDG2024}. This new variable is determined by minimising the function
\begin{equation}
  \chi^2(\lambda) = 1 - \hat{p}_{\Lz}(\lambda) \cdot \hat{f},  
\end{equation}
where $\hat{f}$ is the $\Lambda$ flight direction unit vector and $\lambda$ is a correction factor applied to the pion momentum: $\vec{p}_\pi^{\text{  Corr}} = \lambda \, \vec{p}_\pi$. The corrected $\Lz$ momentum is then given by $\vec{p}_{\Lz} = \vec{p}_\proton + \vec{p}_\pi^{\text{  Corr}}$, where $\hat{p}_{\Lz}(\lambda)$ is its unit vector. The value of $\lambda$ that minimizes the $\chi^2$ is obtained using the \texttt{Minuit} package~\cite{py:iminuit}, and is subsequently used to compute $m_{\text{Corr}}(p\pi)$, assuming the proton–pion mass hypotheses.

The construction of these variables relies only on kinematic constraints and decay topology, and their distributions are found to be insensitive to variations of the assumed signal form factors within current theoretical uncertainties.


\section{Candidate selection}
\label{sec:CandidatesSelection}

The preselections require final state particles with a good track-fit $\chi^2$, that are consistent with the decay topology of both the \decay{\Lz}{\proton \mun \neumb}  and \decay{\Lz}{\proton \pim} decays. The reconstructed $\Lambda$ candidate is required to be significantly displaced from any PV, ensured by requesting a flight-distance–significance requirement greater than 50. Particle-identification requirements are applied to suppress backgrounds from misidentified hadrons and muons. Proton candidates must be well identified and be incompatible with either the kaon or muon hypotheses. Similarly, the companion pion or muon must be incompatible with the kaon identification hypothesis. For the signal channel, the combined proton and muon PID selection has an efficiency of approximately 47\%, while for the normalisation channel the corresponding proton and pion PID efficiency is about 43\%. The proton and the companion track must originate from a common vertex with good fit quality $\chi^2$, and the reconstructed $\Lz$ candidate is required to have a decay time greater than 9\ps, a visible mass below 1141\mevcc, and a reconstructed momentum vector and a reconstructed momentum vector that is only loosely constrained with respect to the PV, reflecting the presence of an undetected neutrino in the final state. This selection retains about 60\% of simulated signal decays.

\begin{figure}[tb]
  \begin{center}

    \includegraphics[width=0.495\linewidth]{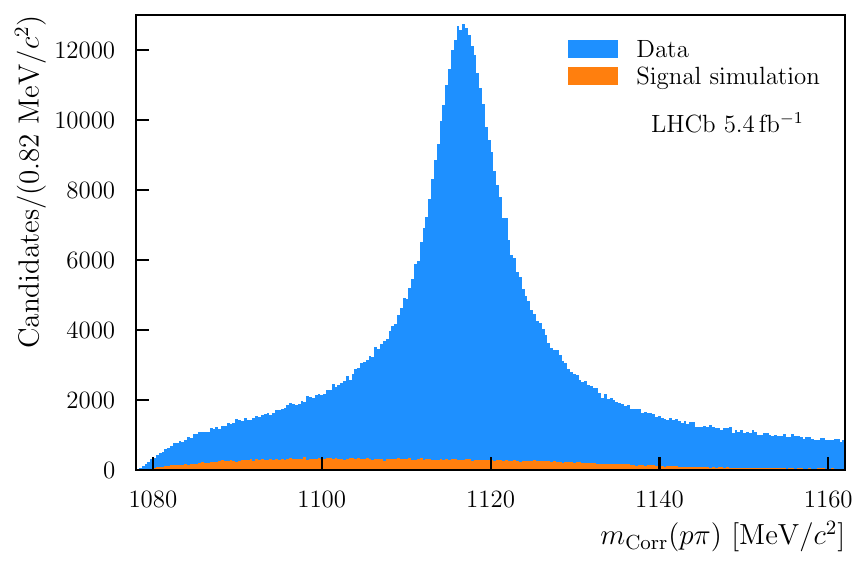}
    \includegraphics[width=0.495\linewidth]{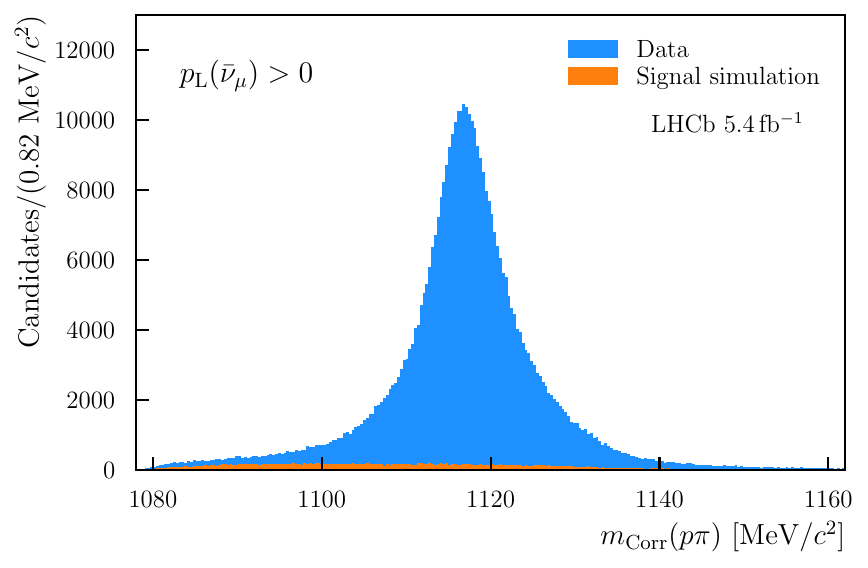}

  \caption{
  Distribution of the $m_{\text{Corr}}(p\pi)$ variable observed in data passing the signal preselection (blue), with the expected signal contribution from simulated \decay{\Lz}{\proton \mun \neumb} decays overlaid (orange), (left) before and (right) after applying the $p_{\rm L}(\neumb) > 0$ requirement, which suppresses combinatorial background.
}
  \label{fig:MppiCorr}
  \end{center}
\end{figure}

The signal selection is developed using simulated samples of \decay{\Lz}{\proton \mun \neumb}, \decay{\Lz}{\proton \pim}  and minimum-bias candidates. Figure~\ref{fig:MppiCorr} shows a substantial suppression of the combinatorial background component, achieved by requiring a positive value of $p_{\rm L}(\neumb)$. The resulting sample is primarily composed of background from \decay{\Lz}{\proton \pim} decays, with a visible signal contribution in the low-side tail of the $m_{\text{Corr}}(p\pi)$ distribution.

\begin{figure}[tb]
  \begin{center}
    \includegraphics[width=0.495\linewidth]{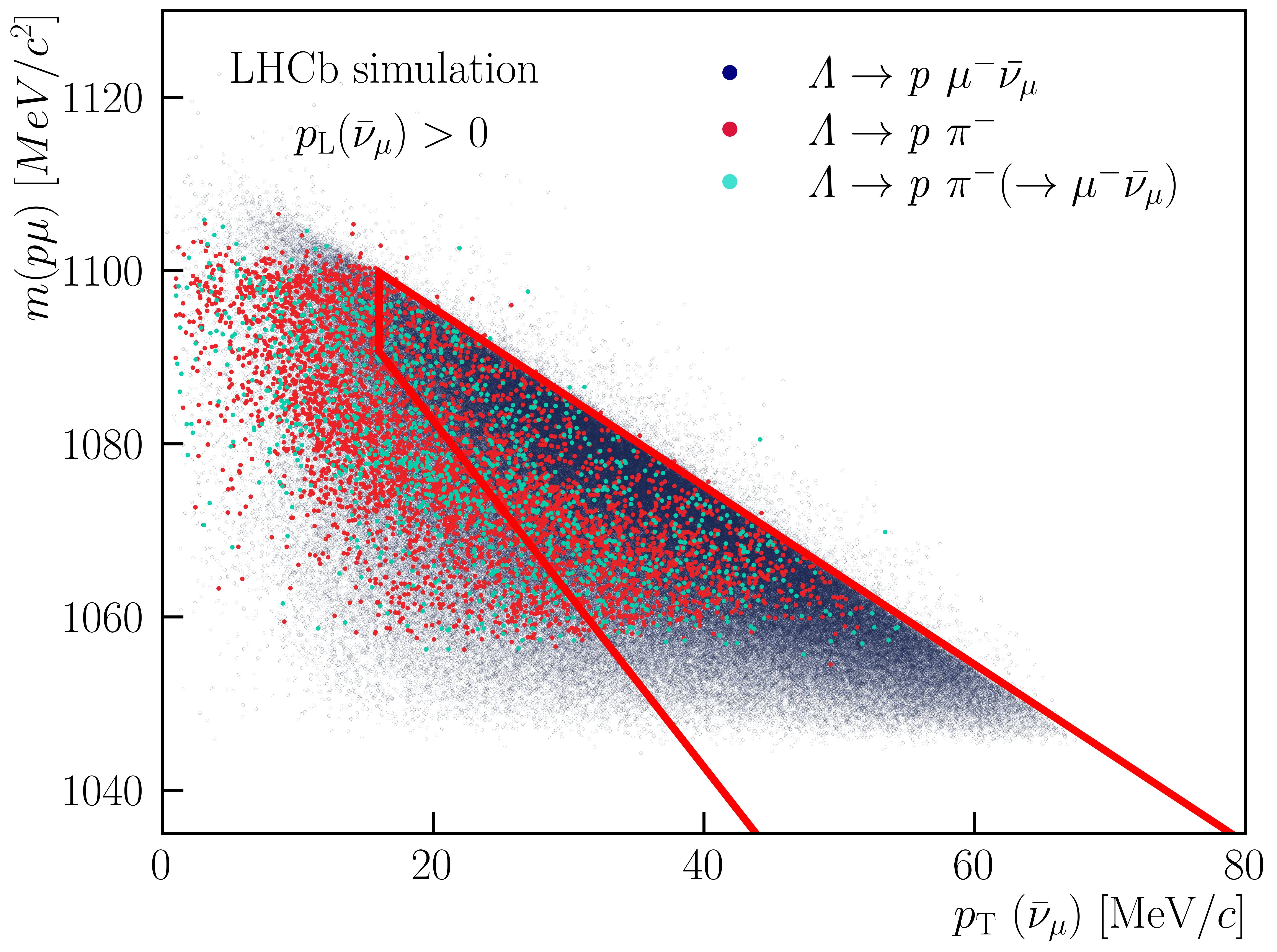}
    \includegraphics[width=0.495\linewidth]{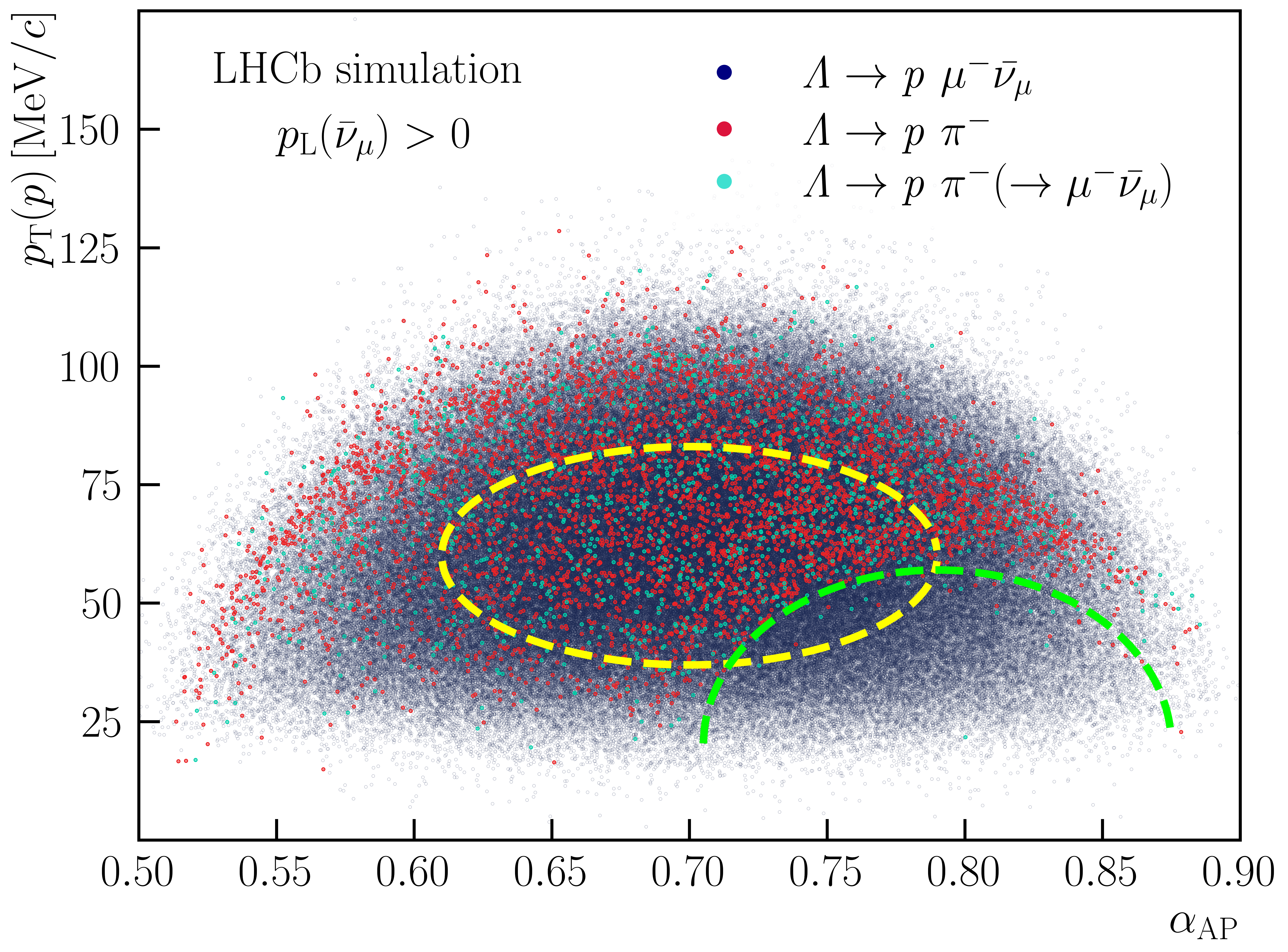}
  \caption{
     Graphical representation of the selection requirements in the (left) $\pt(\neumb)$ vs. $m(p\muon)$ plane and (right) the Armenteros--Podolanski plane. Only candidates inside the red box and inside the yellow ellipse or below the green curve are accepted.}
  \label{fig:Sel1Sel2}
  \end{center}
\end{figure}

Requiring a positive argument inside the square root in the expression for $p_{\rm L}(\neumb)$ imposes a constraint in the $\pt(\neumb)$ vs.\ $m(p\mu)$ plane. Consequently, a tighter selection is applied in this plane (see Fig.~\ref{fig:Sel1Sel2} (left)). In addition, a requirement in the Armenteros--Podolanski~\cite{armenteros} plane (\pt of the oppositely charged decay products with respect to the $\Lz$ direction of flight vs the longitudinal momentum asymmetry $\alpha_{\rm AP} = (p_{\rm L}^+ - p_{\rm L}^-)/(p_{\rm L}^+ + p_{\rm L}^-)$) is introduced, selecting candidates in Fig.~\ref{fig:Sel1Sel2} (right) within the yellow ellipse or below the green curve, where the signal density is higher. An additional requirement of $m_{\text{Corr}}(p\pi) < 1160 \mevcc$ and $m(p\pi)< 1120 \mevcc$, where no signal is expected, is imposed. Regarding the normalisation channel, a requirement is imposed in the Armenteros--Podolanski plane to remove the \decay{\KS}{\pip \pim}  contribution.

\section{Normalisation and signal yield}
\begin{figure}[tb]
  \begin{center}

    \includegraphics[width=0.495\linewidth]{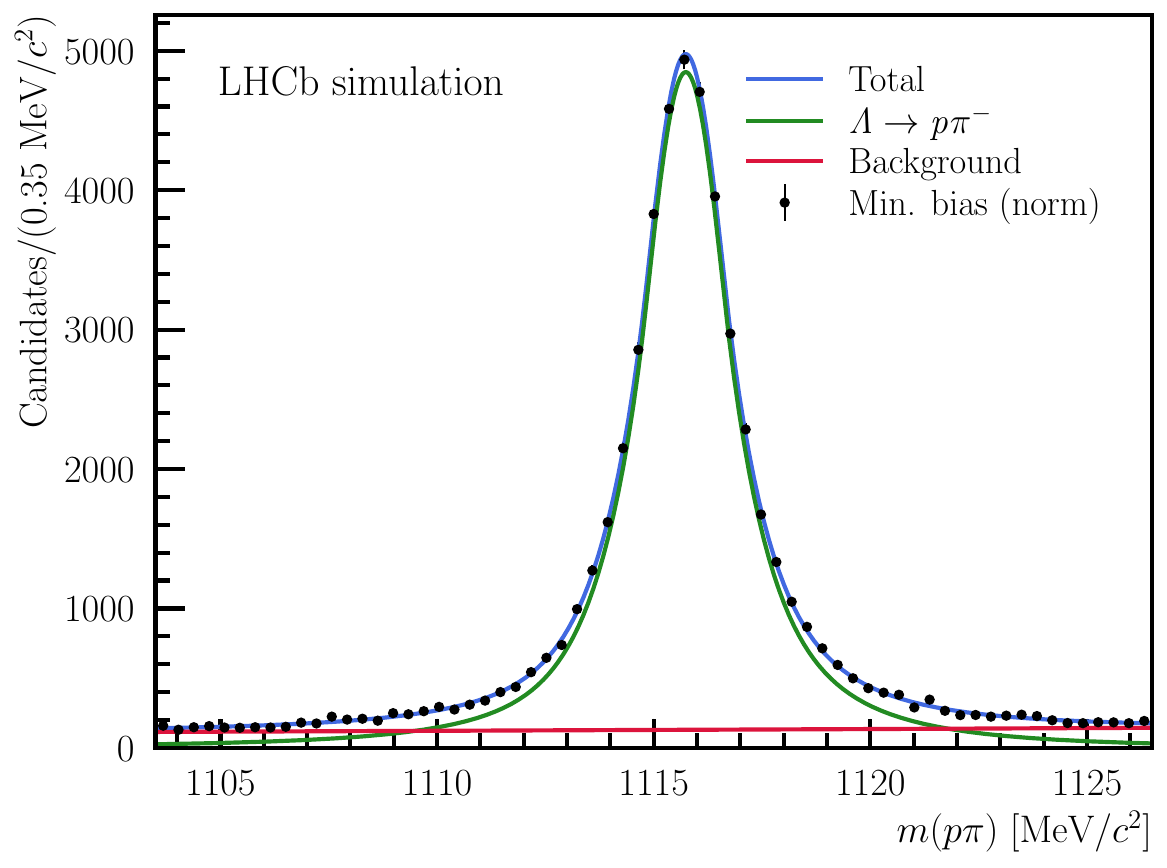}
    \includegraphics[width=0.495\linewidth]{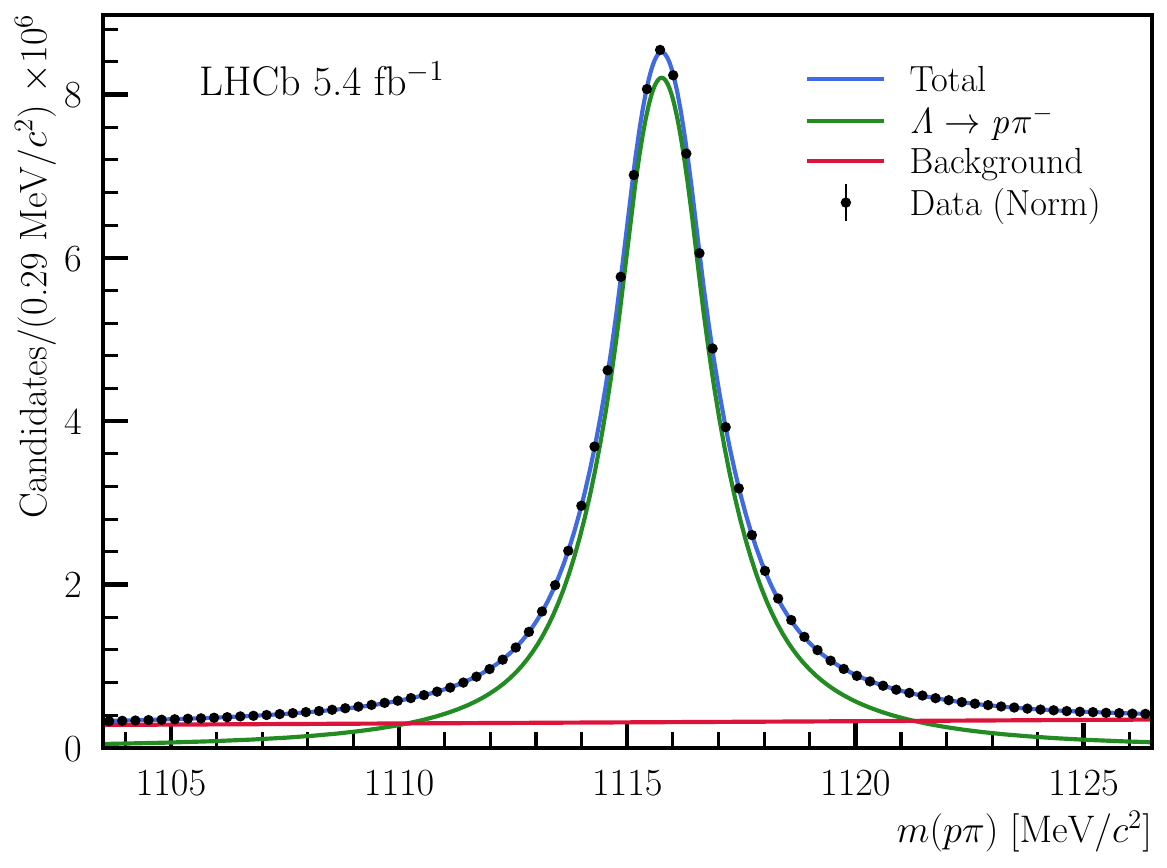}
  \caption{
    Invariant-mass distributions of $m(p\pi)$ for (left) a minimum-bias simulation sample and (right) data candidates passing the normalisation channel selection.}
  \label{fig:FitNormData1}
  \end{center}
\end{figure}

The normalisation yield is obtained from a fit to the $m(p\pi)$ distribution of candidates in data that pass the selection criteria for the normalisation channel. To model the \decay{\Lz}{\proton \pim} shape, simulation is used to obtain a description of the $m(p \pi)$ distribution of the \decay{\Lz}{\proton \pim} decays. The model consists of the sum of a Crystal Ball function~\cite{Skwarnicki:1986xj}, a modified Gaussian function which includes tails on both sides of the peak, and a Johnson~$\mathrm{S_U}$ distribution~\cite{JohnsonSU}. The parameters are determined from an extended unbinned maximum-likelihood fit to the mass distribution of the selected simulated candidates. The tail parameters of the Crystal Ball and Johnson~$\mathrm{S_U}$ functions are constrained with Gaussian priors to the result of the \decay{\Lz}{\proton \pim}  simulation fit, while the peak position and the core width are allowed to vary freely in the fits to data, to account for known residual data-simulation differences. The robustness of the normalisation yield with respect to these constraints is validated by repeating the fit without applying any constraints on the tail parameters. The fit to data, shown in Fig.~\ref{fig:FitNormData1} (right), determines a total normalisation yield of $N_{\Lz \to \proton \pim } = (9.9337 \pm 0.0018) \times 10^{7}$, with the background modelled by an exponential function.

\begin{figure}[bt]
  \begin{center}

    \includegraphics[width=0.495\linewidth, height=5.5cm]{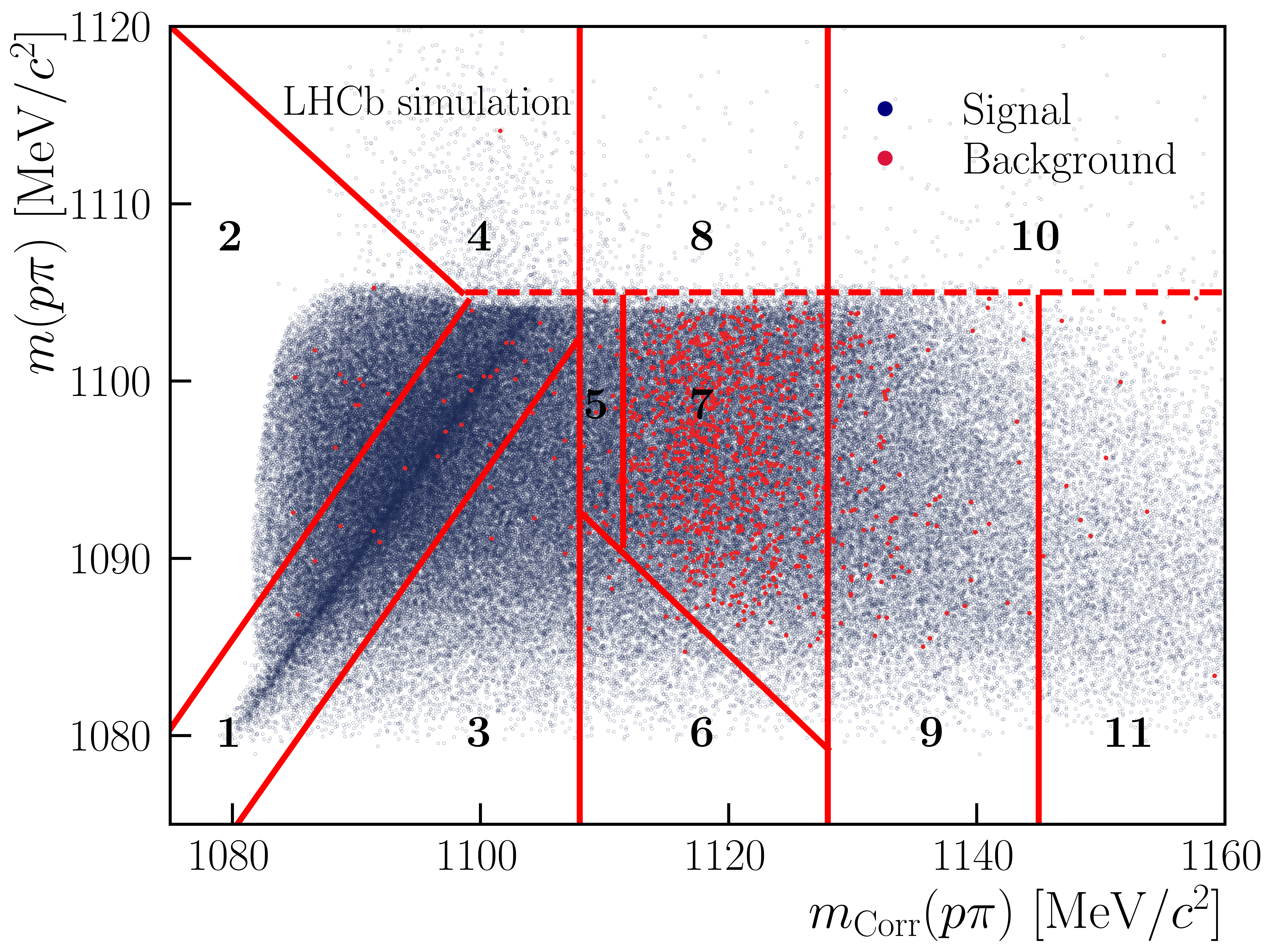}
    \includegraphics[width=0.495\linewidth, height=5.5cm]{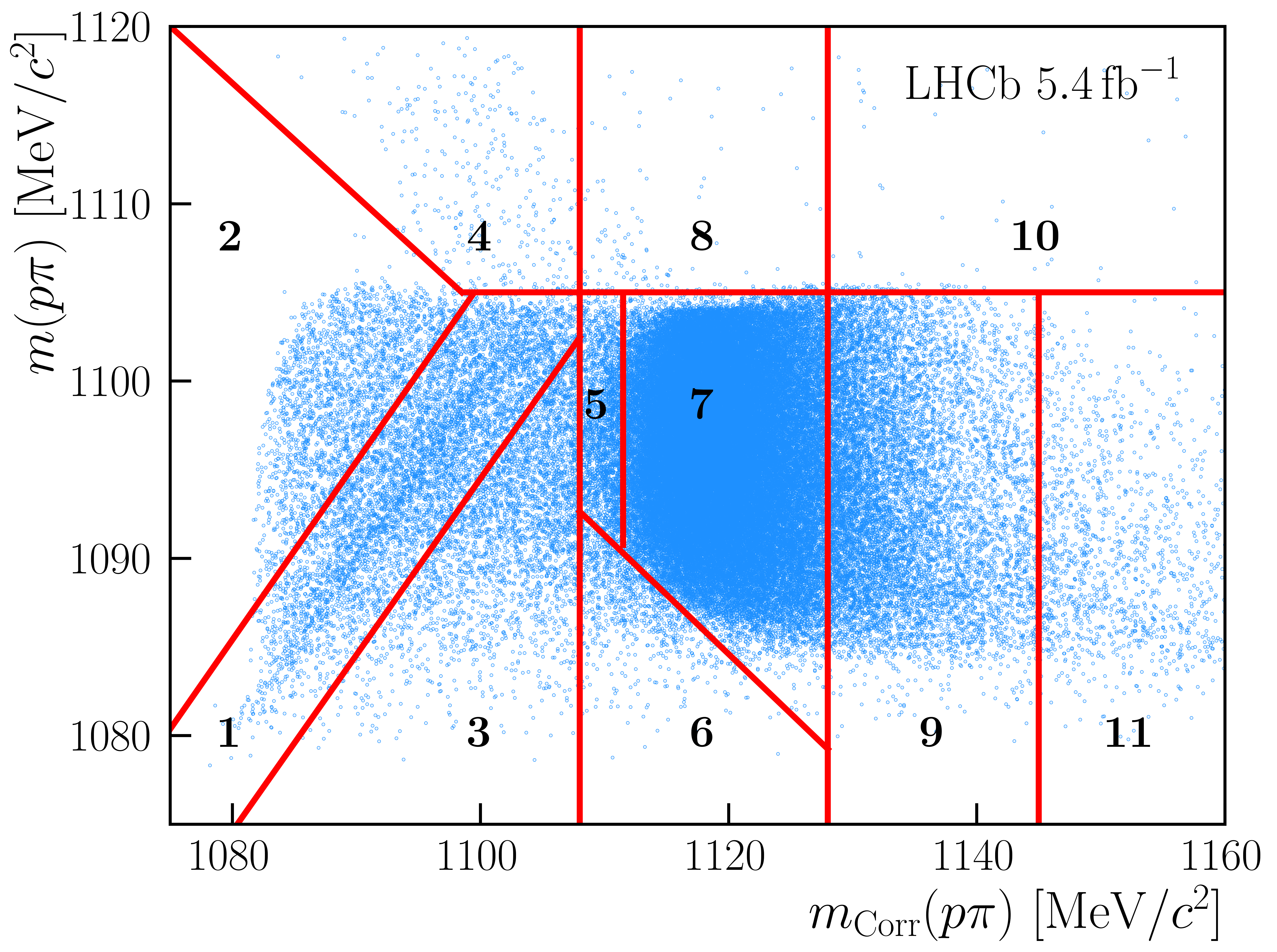}

  \caption{
     Binning scheme of the $m_{\text{Corr}}$(p$\pi$) vs. $m(p\pi)$ plane used to perform a binned fit. Left: distribution of simulated $\decay{\Lz}{\proton \mun \neumb} $ signal and \decay{\Lz}{\proton \pim}  background candidates in the 2D plane. Right: distribution of selected data candidates in the same plane.} 
  \label{fig:BinningFit}
  \end{center}
\end{figure}

The signal yield is measured using a binned maximum-likelihood fit, following the procedure described in Ref.~\cite{CowanStat}, in bins of the $m_{\text{Corr}}(p\pi)$ vs. $m(p\pi)$ plane, with the binning shown in Fig.~\ref{fig:BinningFit}. The fitter uses the observed bin counts as input, with the expectation determined from simulated histograms, and returns the fitted yields for each contribution. A Poisson-based log-likelihood is used to account for limited sample size. For each bin, the fitter computes the quantity
$\chi^{2}$, defined as
\begin{equation}
\chi^{2}
= 2\left(\hat{N} - N_{\text{obs}}
+ N_{\text{obs}} \log\frac{N_{\text{obs}}}{\hat{N}}\right),
\end{equation}
where $N_{\text{obs}}$ is the observed number of selected events in the bin and
$\hat{N}$ is the expected sum of all components in that bin, obtained from the simulation templates. The latter is computed using the fraction of each component in the bin and the overall yield of each component as determined by the fit.

The binning scheme consists of 11 bins in the two-dimensional $m_{\text{Corr}}(p\pi)$ vs.\ $m(p\pi)$ plane. The signal is primarily concentrated along the diagonal where $m_{\text{Corr}}(p\pi) = m(p\pi)$. Bin~1, located on this diagonal, is expected to have the highest signal purity. The adjacent bins to the left (Bin~2) and right (Bin~3) also exhibit high signal purity. These three bins contribute the most to the sensitivity for determining the signal yield in data. Bin~7, located near the centre of the plane, is dominated by background and is intentionally designed to cover the region with the most concentrated background. Bins~5 and~9, adjacent to this central region, contain a significant mixture of both signal and background, which helps to constrain the relative contributions in the fit. The remaining bins serve as control regions, contributing to the overall robustness of the fit.

The fit is performed using simulation templates for \decay{\Lz}{\proton \mun \neumb}, \decay{\Lz}{\proton \pim}, \decay{\Lz}{\proton \pim(\decay{}{\mun \neumb})}, and the combinatorial background. The latter is modelled with candidates from the minimum-bias sample that pass the full signal selection but do not match any of the decays listed above. The resulting signal yield is \mbox{$N_{\Lz \to p\muon \neumb}$ = $(1.637 \pm 0.016)\times 10^4$}, where the uncertainty  accounts for both the statistical fluctuations and the uncertainty in the simulation templates, which is accounted for by repeating the fit 10\,000 times using Poisson-fluctuated templates, and taking the standard deviation of the fitted yields as the uncertainty.

As a cross-check, a machine-learning algorithm is used for finding the best binning-scheme to perform the binned fit. Specifically, a \texttt{DecisionTreeClassifier}~\cite{Scikit-learn-paper}  is used to identify a new binning scheme with 10 bins. The fit is then performed using this new binning, following the same procedure as for the baseline binning scheme. The result is compatible within $0.5 \sigma$ of the baseline binning scheme value.

\begin{figure}[tb]
  \begin{center}

    \includegraphics[width=0.495\linewidth]{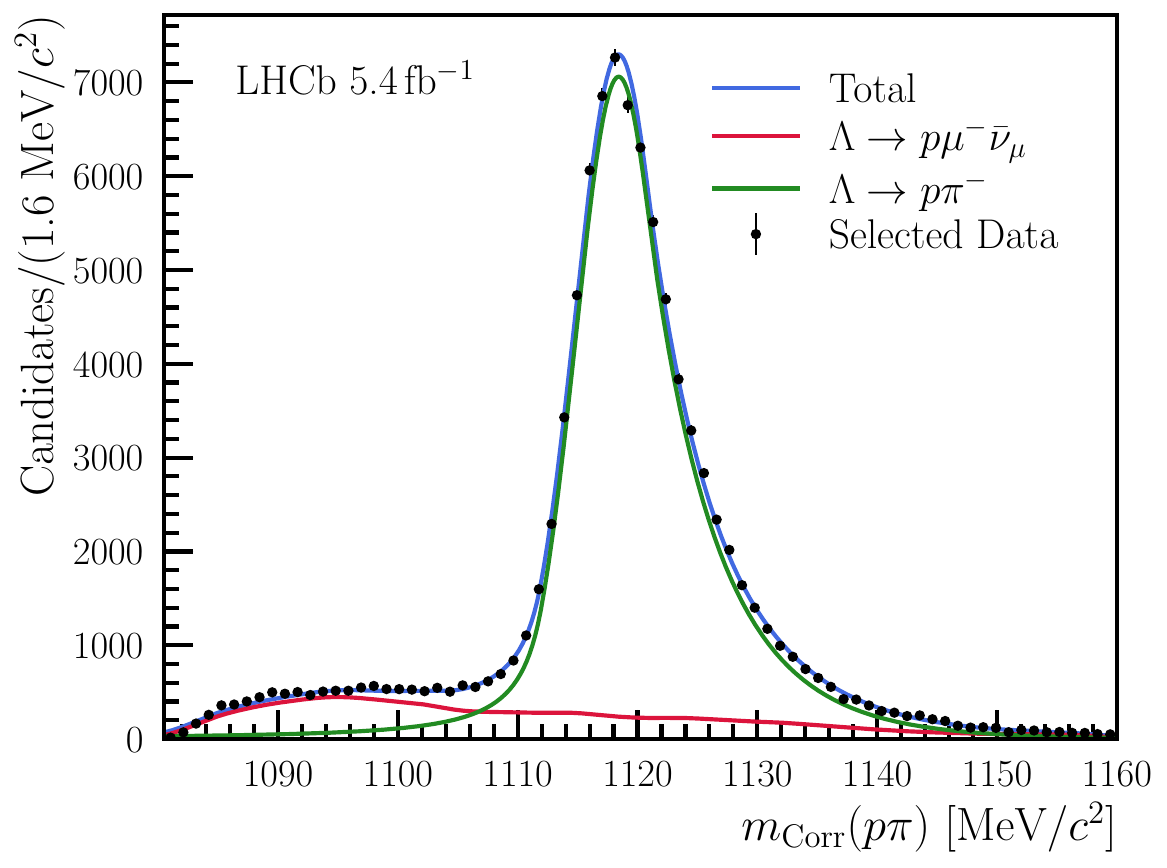}
    
  \caption{Distribution of $m_{\text{Corr}}(p\pi)$ for the selected data, integrated over all bins, with the result of the one-dimensional fit also shown.
     }
  \label{fig:SignalFit}
  \end{center}
\end{figure}

Another cross-check is performed using a one-dimensional fit in $m_{\text{Corr}}(p\pi)$ (Fig.~\ref{fig:SignalFit}), despite the limited size of the combinatorial background simulation, which prevents a fully robust fit. This is possible because, although the limited simulated background statistics prevent the use of a non-parametric modelling such as a kernel density estimate, the \decay{\Lz}{\proton \pim} and \decay{\Lz}{\proton \pim(\decay{}{\mun \neumb})} background components exhibit a smooth peaking behaviour that is well described by a double-sided Crystal Ball function. The signal yield measured by performing this fit, where the signal component is modelled using a Kernel Density Estimation (KDE), is compatible within $0.3 \sigma$ of the two-dimensional fit result. 
As a final check, the one-dimensional fit to the $m_{\text{Corr}}$(p$\pi$) distribution is repeated in four bins of $\pt(\Lz)$ using the same model as described before. The results in each of the four $p_T$ bins are compatible with the central value of the baseline fit within $1\sigma$.

\section{Efficiency determination}

Two key ingredients for the measurement of  the branching fraction $\mathcal{B}(\decay{\Lz}{\proton \mun \neumb} )$ are the observed signal yield and the signal efficiency, defined as the product of efficiencies from event generation, reconstruction, preselection, and the selection detailed in Sec.~\ref{sec:CandidatesSelection}. The signal yield is extracted from a binned maximum-likelihood fit to the full data sample. The total signal efficiency, determined from simulated signal decays, is found to be \mbox{$\epsilon_{\Lz \to p\muon\neumb}$ = $(1.0542  \pm 0.0026) \times 10^{-4}$.}

The efficiency of the normalisation channel is computed by fitting the $m(p\pi)$ distribution in a minimum-bias simulation sample that passes the selection criteria for the normalisation channel, following exactly the same procedure as that used to extract the normalisation yield (see Fig.~\ref{fig:FitNormData1} left). The fitted yield is then divided by the total number of generated \decay{\Lz}{\proton \pim} decays in the simulation sample, resulting in an efficiency of \mbox{$\epsilon_{\Lz \to \proton \pim} = (1.5749 \pm 0.0067) \times 10^{-4}$}.

Combining the signal yield and efficiency with those of the \decay{\Lz}{\proton \pim} channel and its branching fraction, the normalisation constant in Eq.~\ref{eq:number1} is found to be $\alpha = (9.640 \pm 0.090) \times 10^{-9}$, where the quoted uncertainty, apart from the contribution from $\mathcal{B}(\decay{\Lz}{\proton \pim})$, is purely statistical and arises from the limited size of the simulated samples and the uncertainty on the fitted normalisation yield.

\subsection{Corrections to simulated samples}
\label{subsec:corrections}

All simulation samples are corrected to account for observed differences in the $\pt(\Lz)$ and $\eta(\Lz)$ distributions relative to data, using the \texttt{GBReweighter} algorithm from the \texttt{hep\_ml} package~\cite{Rogozhnikov:2016bdp} and the efficiencies are updated. As a signal proxy, \decay{\Lz}{\proton \pim} candidates from data passing the normalisation selection are used, where the background contribution is subtracted using the \textit{sPlot} technique~\cite{Pivk:2004ty} using the $m(\proton \pi)$ invariant mass as discriminating variable.

The normalisation constant is further corrected for known data-simulation discrepancies in particle-identification (PID), tracking, and trigger efficiencies affecting both signal and normalisation channels. These corrections are evaluated using calibration samples and dedicated tools. For PID and tracking, correction factors are computed with \texttt{PIDCalib2}~\cite{LHCb-PUB-2016-021} and \texttt{TrackCalib2}~\cite{LHCb-DP-2013-002} tools, in bins of final-state particle's momentum and pseudorapidity. The trigger correction accounts for differences in the fraction of TIS events between the signal and normalisation channels. It is obtained by measuring the TIS fraction in simulated signal and normalisation channels passing their corresponding  preselections and computing the ratio. The resulting correction factors, which are applied multiplicatively to the normalisation constant, are $\mathcal{C}_{\text{PID}} = 0.897 \pm 0.031$, $\mathcal{C}_{\text{Tracking}} = 0.995 \pm 0.016$, and $\mathcal{C}_{\text{TIS}} = 1.038 \pm 0.031$,  where the quoted uncertainties arise from the limited size of the simulated samples and from the systematic uncertainty associated with the binning used to compute the corrections.

 The final corrected normalisation constant is

\begin{equation*}
    \label{eq:number_alpha_Corr}
    \alpha = (8.928 \pm 0.044 \ \stat \pm 0.436 \ \syst \pm 0.007 \ (\text{norm})) \times 10^{-9},
\end{equation*}
where the first uncertainty combines the different statistical sources, the second uncertainty accounts for systematic effects described in Sec. \ref{sec:Systematics} and the last arises from the branching fraction of the normalisation channel.


\section{Systematic uncertainties}
\label{sec:Systematics}
The entire analysis is structured with the primary goal of minimising systematic uncertainties. It employs data candidates in which the trigger decision was due to other particles produced in the $pp$ collision and a common preselection is used for both modes, differing only in the particle-identification criteria applied to muons and pions. Furthermore, these requirements have been minimised through a kinematics-based selection. As a result, this approach leads to the cancellation of most systematic uncertainties.

Since \decay{\Lz}{\proton \pim} is used as the normalisation channel, its branching ratio uncertainty is included as an independent systematic uncertainty. For the particle-identification and tracking efficiency corrections, the corresponding software tools rely on binning in kinematic and detector occupancy variables. To account for the associated systematic uncertainty, the variation in the corrected efficiencies under different binning schemes is evaluated.

The particle-identification-efficiency correction introduces two sources of uncertainty. The first is statistical, arising from the finite size of the samples used to determine the correction factor (see Sec.~\ref{subsec:corrections}), which accounts for both the signal and normalisation efficiencies, and carries an uncertainty of $3.5\%$. The second originates from the choice of binning scheme in transverse momentum, pseudorapidity, and detector occupancy used to compute the correction. The associated systematic uncertainty is evaluated, for both signal and normalisation, as the maximum deviation from the average value among 14 alternative binning schemes, yielding $1.6\%$ for the signal and $1.0\%$ for the normalisation channel.

The tracking efficiency correction factor also has a statistical component arising from the finite size of the samples used to determine it, corresponding to an uncertainty of $1.6\%$. Regarding the intrinsic tracking uncertainty, since the proton tracks are common to both the signal and normalisation channels, their uncertainties are considered to be highly correlated and thus cancel in the ratio. A tracking uncertainty of $0.8\%$ is assigned to the muon and pion tracks individually, and an additional intrinsic uncertainty of $1.4\%$ is assigned to the pion track due to hadronic interactions. These tracking efficiency uncertainties are assigned following the \lhcb prescription of Ref.~\cite{LHCb-DP-2013-002}. The total uncorrelated tracking uncertainty is therefore $1.8\%$, computed as $\sqrt{(0.8\%)^2 + (0.8\%)^2 + (1.4\%)^2}$.

The systematic uncertainty associated with the determination of the signal yield is the dominant source, and is evaluated by considering four alternative binning schemes and background model configurations. Some binning schemes feature more uniform bin size, while one includes a narrower diagonal region to better capture the signal concentration around the $m_{\text{Corr}}(p\pi) = m(p\pi)$ line. Regarding background modelling, the combinatorial background template is dominated by misidentified \decay{\Lz}{\proton \pim} decays, due to the strong suppression of true combinatorial events. This can reduce the precision of the fit, as part of the \decay{\Lz}{\proton \pim} contribution may be absorbed into the combinatorial component, which is modelled using a simulated sample of significantly smaller size than the \decay{\Lz}{\proton \pim} template. To explore the impact of these modelling choices, the fit is repeated under two alternative configurations in which no separate combinatorial background template is used: one in which the relative contribution of \decay{\Lz}{\proton \pim} and \decay{\Lz}{\proton \pim(\decay{}{\mun \neumb})} background is fixed to the value observed in simulation after selection, and one in which this ratio is allowed to float. 

The systematic uncertainty on the signal yield due to the fit is evaluated from the variation in results obtained using alternative binning schemes and fit models, and is quantified as 3.9\%.

Finally, it is assumed that the efficiency of the trigger selection is independent of the candidate and therefore equal for signal and normalisation candidates passing the preselection. From first principles, this should be the case, as in both scenarios, the parent particle is a $\Lz$ baryon producing two tracks, and the kinematics of the decay are nearly identical. This assumption is validated using simulation. The comparison, although limited by statistical precision, yields a ratio compatible with unity. Nevertheless, given the sizeable uncertainty, the central value of this ratio is used to correct the total efficiency (see Sec.~\ref{subsec:corrections}), with the error on the ratio propagated as the uncertainty.

In simulation, about 3.5 \% of signal candidates are misclassified when applying the truth-based selection. This fraction is estimated by fitting the distribution of candidates labelled as combinatorial background in the signal sample, using templates for signal, \decay{\Lz}{\proton \pim}, and combinatorial decays constructed using KDEs from simulation. To account for possible differences in the selection efficiency between these misidentified and correctly identified signal candidates, a systematic uncertainty is assigned based on their respective efficiencies and relative fractions. The resulting uncertainty is 0.69\%.

Table~\ref{tab:Systematics} summarises the systematic uncertainties. Assuming all sources are uncorrelated, they are combined in quadrature, yielding a total relative uncertainty of $6.8\%$ on the branching fraction.

\begin{table}[bt]
\caption{Systematic uncertainties that affect the $\mathcal{B}(\decay{\Lz}{\proton \mun \neumb})$ measurement. 
}
\centering
\begin{tabular}{c|c}
\hline
\textbf{Source} & \textbf{Relative Uncertainty [\%]} \\
\hline
$\mathcal{B}(\decay{\Lz}{\proton \pim})$ & 0.8 \\
Correction PID efficiencies  & 3.5\\
Binning PID signal correction & 1.6 \\
Binning PID norm. correction & 1.0\\
Correction tracking efficiencies  & 1.6\\
Intrinsic tracking uncertainty &  1.8 \\
$\decay{\Lz}{\proton \mun \neumb}$ yield (fit binning) & 3.9\\
Correction trigger efficiencies & 3.0\\
Simulation truth matching & 0.7 \\
\hline
Total &  6.8\\
\end{tabular}
\label{tab:Systematics}
\end{table}


\section{Results and conclusions}
\label{sec:conclusions}

This analysis uses data collected by the \lhcb experiment from proton-proton collisions at a centre-of-mass energy of 13\,TeV during Run~2 of the LHC (2016–2018) to measure the branching fraction of the $\decay{\Lz}{\proton \mun \neumb}$ decay. To avoid experimenter’s bias, the results were not examined until the full analysis procedure had been finalised. The measurement presented in this analysis yields
\[
\begin{aligned}
\mathcal{B}(\decay{\Lz}{\proton \mun \neumb}) 
&= (1.462 \pm 0.016 \stat \pm 0.100 \syst \pm 0.011 (\text{norm})) \times 10^{-4} \\
&= (1.46 \pm 0.10) \times 10^{-4},
\end{aligned}
\]
corresponding to a total uncertainty of $6.9\%$.

Using the measured branching fraction $\mathcal{B}(\decay{\Lz}{\proton \mun \neumb})$ together with the $\Lz$ lifetime, $\tau_{\Lambda} = (2.617 \pm 0.010) \times 10^{-10}\mathrm{s}$\cite{PDG2024}, the partial decay rate $\Gamma(\decay{\Lz}{\proton \mun \neumb})$ is obtained. The CKM matrix element $|V_{us}|$ is then extracted using lattice QCD form-factor predictions, yielding $|V_{us}| = 0.235 \pm 0.016$ with the more conservative input (Eq.\eqref{eq:LQCD-lattice}) and \mbox{$|V_{us}| = 0.2459 \pm 0.0085$} with the alternative prediction featuring smaller uncertainties but less conservative assumptions (Eq.\eqref{eq:LQCD-lattice2}).

Although these determinations carry larger uncertainties than those derived from kaon decays, they provide an independent constraint that is relevant in light of the long-standing tension in global $|V_{us}|$ fits. Inserting these values into the CKM unitarity relation,
\[
\sqrt{|V_{ud}|^2 + |V_{us}|^2 + |V_{ub}|^2} =
\begin{cases}
    1.0016 \pm 0.0038 & \text{(conservative)}, \\
    1.0042 \pm 0.0021& \text{(smaller uncertainties)},
\end{cases}
\]
yields results consistent with unitarity of the first row of the CKM matrix.

This measurement enables a test of lepton flavour universality in \decay{\squark}{\uquark} quark transitions, where any observed deviation from the expectation would constitute evidence for physics beyond the SM. Using the precisely measured electron-mode branching fraction, $\mathcal{B}(\decay{\Lz}{\proton \electron^- \neueb}) = (8.34 \pm 0.14) \times 10^{-4}$\cite{PDG2024}, the experimental LFU ratio is determined as $R_{\mathrm{exp}}^{\mu e} = 0.175 \pm 0.012$. This value is consistent with the main lattice QCD prediction of $0.1735 \pm 0.0098$\cite{LQCD} at the $0.1\sigma$ level, with the NLO prediction of $0.153 \pm 0.008$\cite{Chang:2014iba} at $1.5\sigma$, and with the more precise but less conservative lattice result of $0.16638 \pm 0.00020$\cite{LQCD} at $0.95\sigma$. The measured ratio is also consistent with recent QCD sum-rule calculations~\cite{wdt2-stt2}, within the quoted uncertainties.

Compared to the previous measurement by the BESIII Collaboration~\cite{BESIII:2021ynj},
which was performed using a double-tag technique in \decay{\jpsi}{\Lz \Lbar} events,
this result achieves a factor of two improvement in precision while remaining
consistent within uncertainties. The improved precision is primarily driven by the
very large $\Lambda$ baryon sample available in proton--proton collisions at the LHC,
which compensates for the more challenging experimental environment. This measurement
therefore provides a complementary and more precise determination of
$\mathcal{B}(\decay{\Lz}{\proton \mun \neumb})$, strengthening its impact on tests of
lepton flavour universality and on the extraction of $|V_{us}|$ from baryonic
semileptonic decays.

The measured LFU ratio is also consistent with the BESIII determination,
$R_{\mathrm{exp}}^{\mu e} = 0.178 \pm 0.028$, with the difference between the two
measurements corresponding to approximately $0.1\sigma$. The improved precision of
the present result, more than a factor of two relative to BESIII, provides the most
stringent test of lepton flavour universality in $\Lz \to \proton \, \ell^- \bar{\nu}_\ell$ decays to date.

%% file: acknowledgements.tex
\section*{Acknowledgements}
%
%
\noindent We express our gratitude to our colleagues in the CERN
accelerator departments for the excellent performance of the LHC. We
thank the technical and administrative staff at the LHCb
institutes.
We acknowledge support from CERN and from the national agencies:
ARC (Australia);
CAPES, CNPq, FAPERJ and FINEP (Brazil); 
MOST and NSFC (China); 
CNRS/IN2P3 (France); 
BMFTR, DFG and MPG (Germany);
INFN (Italy); 
NWO (Netherlands); 
MNiSW and NCN (Poland); 
MCID/IFA (Romania); 
MICIU and AEI (Spain);
SNSF and SER (Switzerland); 
NASU (Ukraine); 
STFC (United Kingdom); 
DOE NP and NSF (USA).
We acknowledge the computing resources that are provided by ARDC (Australia), 
CBPF (Brazil),
CERN, 
IHEP and LZU (China),
IN2P3 (France), 
KIT and DESY (Germany), 
INFN (Italy), 
SURF (Netherlands),
Polish WLCG (Poland),
IFIN-HH (Romania), 
PIC (Spain), CSCS (Switzerland), 
and GridPP (United Kingdom).
We are indebted to the communities behind the multiple open-source
software packages on which we depend.
Individual groups or members have received support from
Key Research Program of Frontier Sciences of CAS, CAS PIFI, CAS CCEPP, 
Fundamental Research Funds for the Central Universities,  and Sci.\ \& Tech.\ Program of Guangzhou (China);
Minciencias (Colombia);
EPLANET, Marie Sk\l{}odowska-Curie Actions, ERC and NextGenerationEU (European Union);
A*MIDEX, ANR, IPhU and Labex P2IO, and R\'{e}gion Auvergne-Rh\^{o}ne-Alpes (France);
Alexander-von-Humboldt Foundation (Germany);
ICSC (Italy); 
Severo Ochoa and Mar\'ia de Maeztu Units of Excellence, GVA, XuntaGal, GENCAT, InTalent-Inditex and Prog.~Atracci\'on Talento CM (Spain);
SRC (Sweden);
the Leverhulme Trust, the Royal Society and UKRI (United Kingdom).

%% file: Authorship_LHCb-PAPER-2025-030.tex
\centerline
{\large\bf LHCb collaboration}
\begin
{flushleft}
\small
R.~Aaij$^{38}$\lhcborcid{0000-0003-0533-1952},
A.S.W.~Abdelmotteleb$^{57}$\lhcborcid{0000-0001-7905-0542},
C.~Abellan~Beteta$^{51}$\lhcborcid{0009-0009-0869-6798},
F.~Abudin{\'e}n$^{57}$\lhcborcid{0000-0002-6737-3528},
T.~Ackernley$^{61}$\lhcborcid{0000-0002-5951-3498},
A. A. ~Adefisoye$^{69}$\lhcborcid{0000-0003-2448-1550},
B.~Adeva$^{47}$\lhcborcid{0000-0001-9756-3712},
M.~Adinolfi$^{55}$\lhcborcid{0000-0002-1326-1264},
P.~Adlarson$^{85}$\lhcborcid{0000-0001-6280-3851},
C.~Agapopoulou$^{14}$\lhcborcid{0000-0002-2368-0147}
% ...

{\footnotesize \it

% ...
$^{14}$Universit{\'e} Paris-Saclay, CNRS/IN2P3, IJCLab, Orsay, France\\
% ...
$^{38}$Nikhef National Institute for Subatomic Physics, Amsterdam, Netherlands\\
% ...
$^{47}$Instituto Galego de F{\'\i}sica de Altas Enerx{\'\i}as (IGFAE), Universidade de Santiago de Compostela, Santiago de Compostela, Spain\\
% ...
$^{51}$Physik-Institut, Universit{\"a}t Z{\"u}rich, Z{\"u}rich, Switzerland\\
% ...
$^{55}$H.H. Wills Physics Laboratory, University of Bristol, Bristol, United Kingdom\\
% ...
$^{57}$Department of Physics, University of Warwick, Coventry, United Kingdom\\
% ...
$^{60}$School of Physics and Astronomy, University of Glasgow, Glasgow, United Kingdom\\
$^{61}$Oliver Lodge Laboratory, University of Liverpool, Liverpool, United Kingdom\\
% ...
$^{69}$Syracuse University, Syracuse, NY, United States\\
% ...
$^{85}$Department of Physics and Astronomy, Uppsala University, Uppsala, Sweden, associated to $^{60}$\\
% ...
}
\end{flushleft}